\documentclass[12pt]{article}
\pdfoutput=1

\usepackage[utf8]{inputenc}
\usepackage[T1]{fontenc}
\usepackage{graphicx}
\usepackage{wrapfig}
\usepackage{epsf}
\usepackage{amsfonts}
\usepackage{color}
\usepackage{mathrsfs}

\definecolor{brown}{cmyk}{0,1,.9,.2}

\def\timesbox{\hbox{$\scriptscriptstyle\times$}}
\def\ant{ {{\lower 1ex  \timesbox} \atop {\raise 1.5ex  \timesbox}}}
\newcommand\ZZ{{\hbox{ Z\kern-1.6mm Z}}}
\newcommand{\Iop}{\relax{\rm I\kern-.18em I}}
\newcommand{\Lop}{\relax{\rm I\kern-.18em L}}
\newcommand{\dop}{\relax{\rm I\kern-.8em d}}
\newcommand{\one}{{\hbox{ 1\kern-1.2mm l}}}

\newcommand{\beq}{\begin{equation}}
\newcommand{\eeq}{\end{equation}}
\newcommand{\bea}{\begin{eqnarray}}
\newcommand{\eea}{\end{eqnarray}}

\newcommand{\lt}{\left}
\newcommand{\rt}{\right}

\newcommand{\del}{\partial}

\newcommand{\al}{\alpha}

\newcommand{\dlt}{\delta}

\newcommand{\s}{\sigma}

\newcommand{\omg}{\omega}

\newcommand{\Dlt}{\Delta}

\newcommand{\Omg}{\Omega}

\newcommand{\Dt}{\tilde D}

\newcommand{\et}{\tilde e}

\newcommand{\gammat}{\tilde \gamma}

\newcommand{\omgt}{\tilde \omega}

\newcommand{\xit}{\tilde \xi}

\newcommand{\delt}{\tilde \partial}

\newcommand{\zb}{\bar z}

\newcommand{\elb}{{\underline e}}

\newcommand{\pilb}{{\underline \pi}}

\newcommand{\Yh}{\hat Y}

\newcommand{\eh}{\hat e}

\newcommand{\yh}{\hat y}
\newcommand{\zh}{\hat z}

\newcommand{\kappah}{\hat \kappa}
\newcommand{\omgh}{\hat \omega}

\newcommand{\delh}{\hat \partial}

\newcommand{\cF}{{\cal F}}

\newcommand{\cL}{{\cal L}}

\newcommand{\cM}{{\cal M}}

\newcommand{\cQ}{{\cal Q}}
\newcommand{\cR}{{\cal R}}
\newcommand{\cS}{{\cal S}}

\newcommand{\cU}{{\cal U}}

\newcommand{\ha}{\hbox{a}}
\newcommand{\hb}{\hbox{b}}
\newcommand{\hc}{\hbox{c}}
\newcommand{\hd}{\hbox{d}}
\newcommand{\he}{\hbox{e}}

\newcommand{\Exp}{{\hbox{Exp}}}

\begin{document}

{}~
{}~
\hfill\vbox{\hbox{IMSc/2014/07/4}}

\vskip 2cm

\centerline{\Large \bf  A cut-off tubular geometry of loop space}

\medskip

\vspace*{4.0ex}

\centerline{\large \rm Partha Mukhopadhyay }

\vspace*{4.0ex}

\centerline{\large \it The Institute of Mathematical Sciences}
\centerline{\large \it C.I.T. Campus, Taramani}
\centerline{\large \it Chennai 600113, India}

\medskip

\centerline{E-mail: parthamu@imsc.res.in}

\vspace*{5.0ex}

\centerline{\bf Abstract}
\bigskip

Motivated by the computation of loop space quantum mechanics as indicated in \cite{semi-classical}, here we seek a better understanding of 
the tubular geometry of loop space $\cL\cM$ corresponding to a Riemannian manifold $\cM$ around the submanifold of vanishing loops. Our 
approach is to first compute the tubular metric of $(\cM^{2N+1})_{C}$ around the diagonal submanifold, where $(\cM^N)_{C}$ is the 
Cartesian product of $N$ copies of $\cM$ with a cyclic ordering. This gives an infinite sequence of tubular metrics such that the one relevant to 
$\cL\cM$ can be obtained by taking the limit $N\to \infty$. Such metrics are computed by adopting an indirect method where the general 
tubular expansion theorem of \cite{tubular} is crucially used. We discuss how the complete reparametrization isometry of loop space arises in 
the large-$N$ limit and verify that the corresponding Killing equation is satisfied to all orders in tubular expansion. These tubular metrics can 
alternatively be interpreted as some natural Riemannian metrics on certain bundles of tangent spaces of $\cM$ which, for $\cM \times \cM$, 
is the tangent bundle $T\cM$.

\newpage

\tableofcontents

\baselineskip=18pt

\section{Introduction and summary}
\label{intro}

Configuration space of a non-linear sigma model (NLSM) \cite{review} with target space $\cM$ is the corresponding free loop space $\cL \cM$ \cite{loop-space}.
\bea
\cL \cM &=& C^{\infty}(S^1, \cM)~.
\eea
One can rewrite NLSM as a single particle model in $\cL \cM$ such that the complete geometry of the
configuration space is manifest \cite{witten82, frenkel, dwv, talks, semi-classical}. For {\it small loops} which are entirely contained in a single
normal neighbourhood \cite{normal} in $\cM$, the configuration space can be described as a tubular neighbourhood \cite{spivak, FS, tubes} around $\Dlt \hookrightarrow \cL \cM$ (see \cite{semi-classical}). Here $\Dlt (\cong \cM)$ is the submanifold of vanishing loops, i.e. constant maps from $S^1$ to $\cM$. It was demonstrated in \cite{semi-classical} that in a semi-classical limit ($\alpha' \to 0$) of the loop space quantum mechanics (LSQM) the string wavefunction localises on $\Dlt$. This in turn implies that the semi-classical expansion is to be derived from tubular expansion of geometric quantities of $\cL \cM$.\footnote{To further elaborate on the relation between standard 2d QFT language of NLSM and LSQM, one notes that for quantization of {\it small strings} in curved space it is crucial that the center of mass (CM) mode and modes of internal fluctuations be separated carefully in a covariant manner. The submanfold structure in $\cL \cM$ gives this description, where $\Delta$ is the space of CM. Given such a separation, the covariant expansion of the string field is given by the tubular expansion of Schr\"odinger wavefunction in LSQM. Simlarly, for the expectation values of operators, one first writes them in position space representation of LSQM and then tubular expands the relevant geometric quantities which include differential operators. } It was also shown that the effective quadratic action for the tachyon state was correctly reproduced at leading order.

There are two obstacles in performing the computation described in \cite{semi-classical},
\begin{enumerate}
\item
Lack of a detailed understanding of the tubular geometry in $\cL\cM$.

\item
$\cL\cM$ is infinite dimensional. This causes divergences to appear in the computation.\footnote{These are analogue of the usual UV divergences of NLSM in the present approach.} 
\end{enumerate} 
To regularise the divergences, it is natural to think of considering a finite dimensional model which, in the limit of large dimensionality, will approach the loop space model. The goal of this work is to construct suitable finite dimensional geometries, perform the required limit and explicitly construct the tubular geometry in $\cL\cM$ in this way. 

As will be discussed in \S \ref{tubnbh}, the finite dimensional {\it cut-off space} we consider is the tubular neighbourhood of the diagonal submanifold $\Dlt$ embedded in $(\cM^{2N+1})_C$, where,
\bea
\cM^N &=& \cM \times \cM \times \cdots \times \cM \quad [\hbox{$N$ factors}]~,
\eea
and the subscript $C$ indicates that the factors are cyclically ordered. The limit of large dimensionality is simply the limit $N\to \infty$. Notice that we use the same notation $\Dlt$ for both the submanifold of vanishing loops in $\cL\cM$ and the diagonal submanifold of $\cM^N$, as both of them are isomorphic to $\cM$ and the embeddings merge together at large $N$.

The plan of the paper is as follows. In \S \ref{tubnbh} we describe the construction of the relevant tubular neighbourhoods. We evaluate the tubular geometry around $\Dlt \hookrightarrow \cM^N$ and the subsequent large-$N$ limit in detail in \S \ref{tub:M^N} and 
\S \ref{tub:LM} respectively with many technical details summarised in several appendices. Finally, in \S \ref{comments} we make comments regarding our choice of  the cut-off space and some possible applications of our results besides its use in the study of LSQM.

Before we end this introduction, below we briefly discuss a couple of key points relevant to our analysis. The first question is: What do we actually mean by ``evaluating tubular geometry" in this case? To answer this we recall that given an arbitrary submanifold admitting a tubular neighbourhood, it is possible to perform covariant Taylor expansion of tensors around the submanifold \cite{tubes, tubular}. The expansion coefficients are tensors of the ambient space evaluated at the submanifold which describe various extrinsic properties of the embedding. In our case both the submanifold and the ambient space are constructed out of $\cM$ only. Therefore, it should be possible to evaluate all the tubular expansion coefficients of, say vielbein, in terms of the intrinsic geometric data of $\cM$ (hereafter called {\it $\cM$-data}). A direct method was used in \cite{semi-classical} by explicitly constructing the Fermi normal coordinates (FNC) \cite{FS, tubes, tubular} and implementing the required coordinate transformation order-by-order. It is difficult to carry out such a method as the computation soon becomes involved enough. In this work we use an indirect method (to be discussed in \S \ref{s:indirect}) which utilizes the general results of \cite{tubular} very crucially and we are able to derive all-order-results this way. 

The second question is: How do we know that the large-$N$-geometry we obtain this way is indeed the geometry of loop space?  We show evidence for the case through arguments involving isometries. In particular, we show that the large-$N$-geometry admits the reparametrization isometry of loop space by verifying the relevant Killing equation to all orders in tubular expansion. We hope to present further evidences with the analysis of geodesics in \cite{cut-offII}.

\section{Construction of tubular neighbourhood}
\label{tubnbh}

Here we describe the construction of tubular neighbourhoods in $\cL \cM$ and $\cM^N$. For $\cL \cM$ we review the intuitive manner in which it was introduced in \cite{semi-classical}. All our discussion about finite dimensional spaces will be rigorous and we would like to think 
of ${\cal L}\cM$ only as a limit.

We will use the notation $\cU_S$ to refer to certain open normal neighbourhood in $S$, where $S$ will stand for $\cM$, $\cM^N$ or $\cL\cM$. We will see how such neighbourhoods are inter-related. A {\it small loop} in $\cM$ is a loop which is entirely contained in a single normal neighbourhood in $\cM$, say $\cU_{\cM}$. All such loop configurations within $\cU_{\cM}$ define a neighbourhood 
$\cU_{\cL\cM}$ in $\cL\cM$ such that $\Dlt \cap \cU_{\cL\cM}$ is non-vanishing. In fact, given that $\Dlt$ is the submanifold of vanishing loops, we must have: $\Dlt \cap \cU_{\cL\cM} \cong \cU_{\cM}$. This simply implies that the points in $\cU_{\cM}$, i.e. zero loops, can be identified with the points in $\Dlt \cap \cU_{\cL\cM}$. Therefore a non-zero loop in $\cU_{\cM}$ corresponds to a point in $\cU_{\cL\cM}$ which is away from the submanifold. Given any point $P$ in a tubular neighbourhood, there exists a unique geodesic passing through $P$ that is orthogonal to the submanifold. The intersection of this geodesic and the submanifold is also a unique point $Q$ \cite{FS}. Since in our case $P\in \cU_{\cL\cM}$ is a loop in $\cU_{\cM}$, one may view the unique point $Q\in \cU_{\cM}$ \footnote{Note that $Q$ is a point in $\Dlt \cap \cU_{\cL\cM}$, but the latter has been identified with $\cU_{\cM}$.} as an average, or in the language of \cite{semi-classical}, centre of mass (CM) of the loop. To define the CM we proceed as follows \cite{semi-classical}. Let $l: S^1 \to \cU_{\cM}$ be the loop corresponding to the point $P$ in $\cU_{\cL \cM}$. Given an arbitrary point $q\in \cU_{\cM}$, we construct the pre-image of $l$ in $T_q\cM$ under the exponential map based at $q$,
\bea
t (q, \s) := \Exp^{-1}_q \circ l(\s)~.
\label{t(q, s)}
\eea 
Since exponential map is a diffeomorphism within a normal neighbourhood, the above map is one-to-one.
CM of $l$ is the unique point $Q$ for which,
\bea
\oint d\s t(Q, \s) = 0~.
\label{oint-t}
\eea

The above prescription explicitly spells out how to identify, in a one-to-one manner, all possible small loop configurations in $\cU_{\cM}$ and 
points in $\cU_{\cL \cM}$ in the following way. First of all, the configurations of all possible small loops in $\cU_{\cM}$ are in one-to-one 
correspondence with certain loop configurations in $T\cM$ such that each loop in $T\cM$ resides entirely in a single fiber with its average
position fixed at the corresponding base point. Zero section of $T\cM$ is then identified with $\Dlt \hookrightarrow \cL\cM$ and the loops in $T_Q\cM$ are 
identified with points on the geodesics in $\cL\cM$ that intersect $\Dlt \hookrightarrow \cL\cM$ orthogonally at $Q\in \Dlt$.\footnote{The
tubular neighbourhood theorem would then demand that the one-to-one map described above be a diffeomorphism and the identification of
loops in $T\cM$ with points on geodesics in $\cL\cM$ be consistent. In this work we will construct this diffeomorphism explicitly for the 
cut-off space while postponing the analysis of geodesics to \cite{cut-offII}. \label{f:diffeo} }

The above discussion however does not directly tell us how to write down the metric of $\cL\cM$ in the form of a tubular expansion in the sense of \cite{tubular}. Explicitly writing down this metric is the purpose of this work. To this end we define an infinite sequence of finite dimensional spaces ${\cal L}^{(N)}\cM$ ($N$ being a positive integer) such that the tubular geometry of ${\cal L}\cM$ can be understood as a large-$N$ limit. 
Such a cut-off loop space should satisfy the following properties,
\begin{enumerate}
\item
$\cL^{(N)}\cM$ admits $\Dlt \cong \cM$ as a submanifold. 

\item
Tubular geometry around $\Dlt \hookrightarrow \cL^{(N)}\cM$ approaches the tubular geometry around $\Dlt \hookrightarrow \cL\cM$ in the limit $N\to \infty$.
\end{enumerate}
As mentioned in \S \ref{intro}, in this work we explore the following possibility,
\bea
\cL^{(N)}\cM &=& (\cM^{2N+1})_C ~,
\label{cut-off}
\eea
with the diagonal submanifold of $\cM^{2N+1}$ playing the role of $\Dlt$. 

The construction of tubular neighbourhood around $\Dlt \hookrightarrow \cM^N$ is simply a discretised version of the above discussion regarding loop space. Since $\cU_{\cM^N} = (\cU_{\cM})^N$, a point in $\cU_{\cM^N}$, say $(l_1, l_2, \cdots ,l_N)$ is an $N$-point configuration in $\cU_{\cM}$. Following the loop-space-discussion we define the average position/CM by the unique point $Q \in \cU_{\cM} (\hbox{or } \Dlt \cap \cU_{\cM^N})$ such that,
\bea
\sum_{p=1}^N t_p(Q) &=& 0~,
\label{sum-tp}
\eea
where given $q\in \cU_{\cM}$,
\bea
t_p(q) &\equiv& \Exp_q^{-1} \circ l_p ~.
\label{tp-lp}
\eea
Notice that it is the condition in (\ref{sum-tp}) which singles out the diagonal submanifold as the space of all possible locations of the CM. 
Therefore all possible $N$-point configurations contained entirely in a single normal neighbourhood in $\cM$ are in one-to-one 
correspondence with certain $N$-point configurations in $T\cM$ such that in each such configuration all the $N$-points are in a single fiber 
with the average fixed at the corresponding base point. Just like in the case of loop space, we then identify the zero section with $\Dlt 
\hookrightarrow \cM^N$ such that the $N$-point configuration in $T_Q\cM$ given by $\{t_p\}$ in (\ref{sum-tp}) is mapped to $(l_1, l_2, 
\cdots , l_N)$ in the neighbourhood of $\Dlt$ in $\cM^N$. This is the basic construction that will be used in \S \ref{tub:M^N} to compute the 
tubular metric in $\cM^N$. 

\section{Tubular geometry in $\cM^N$}
\label{tub:M^N}

The purpose of this section is to explicitly work out the tubular geometry around $\Dlt \hookrightarrow \cM^N$ in terms of 
$\cM$-data. Here we first pose the problem in more technical terms. It is a well-known fact in Riemannian geometry \cite{petersen} that an open neighbourhood around the diagonal of $\cM \times \cM$ is diffeomorphic to an open neighbourhood around the zero section of $T\cM$. The construction described in the previous section is a generalisation of the same statement where $\cM \times \cM$ is replaced by $\cM^N$ and $T\cM$ by a bundle $T^{(N-1)}\cM$ whose base is $\cM$ and the fiber at $Q\in \cM$ is given by\footnote{The reason that there are 
$(N-1)$ additive factors in eq.(\ref{TQ}) is simply because in the $T\cM$ description, as discussed below eq.(\ref{tp-lp}), the $N$-point configuration is given by $(N-1)$ independent tangent vectors of $\cM$. },
\bea
T_Q^{(N-1)}\cM &=& T_Q \cM \oplus \cdots \oplus T_Q\cM~, \quad [(N-1)\hbox{ additive factors}] ~.
\label{TQ}
\eea

The relevant diffeomorphism may be considered to be a transformation between the coordinate systems which are natural in $\cM^N$ and $T^{(N-1)}\cM$. In the 
natural coordinate system in $\cM^N$, hereafter to be called {\it direct product coordinates} (DPC), the point $(l_1, l_2, \cdots , l_N) \in {\cal U}_{\cM}$ is given by,
\bea
\zb^{\bar a} &=& (x_1^{\alpha_1}, x_2^{\alpha_2}, \cdots , x_N^{\alpha_N}) ~.
\label{z-bar-gen}
\eea
On the other hand the natural coordinate system in $T^{(N-1)}\cM$ is taken to be the FNC relevant to the present case. The notation for FNC in a generic case is already set up in Appendix \ref{a:tubular}. See, for example, eq.(\ref{FNC}). In our special case, because of the particular structure of eq.(\ref{TQ}), this takes the following form,
\bea
\hat z^a &=& (x^{\al}, \yh^A) = (x^{\alpha}, \yh_1^{\hat \alpha_1}, \cdots , \yh_{N-1}^{\hat \alpha_{N-1}} ) ~,
\label{zhat}
\eea
where $x^{\al}$ is a general coordinate\footnote{We will eventually take $x^{\al}$ and $x_p^{\al_p}$ ($p=1, 2, \cdots , N$) to be the same local coordinate system in $\cM$.} for $Q\in \cM$ and $\hat y_{\ha}$ ($\ha = 1, 2, \cdots , (N-1)$) is the fiber coordinate along the $\ha$-th factor on the RHS of (\ref{TQ}).

The metric is of course known in DPC in terms of $\cM$-data. A direct method of computing the tubular metric (i.e. the metric in $\hat z$-system given in tubular expansion form) will be to start with the metric in DPC and then perform the coordinate transformation $\bar z \to \hat z$. We will explicitly construct the full coordinate transformation in Appendix \ref{a:coord-construction} in terms of exponential maps. However, as mentioned earlier, this computation is very cumbersome. We will instead adopt an indirect method using which we are able to compute all-order results for vielbein.

As an interesting aside, notice that the above construction naturally gives a Riemannian metric on $T^{(N-1)}\cM$. This Riemannian metric is nothing but the tubular metric that we have set out to compute. For $N=2$ the bundle under question is the tangent bundle $T\cM$ and the explicit form of its metric will be written down to quartic order in eqs.(\ref{g-al-beta-2}-\ref{g-A-B-2}). There are other approaches of constructing natural Riemannian metrics on $T\cM$ in the literature \cite{TM} and it may be interesting to explore if there is any relation between these two types of constructions. 

The indirect method will be described in \S \ref{s:indirect}. The results for vielbein-expansion will be obtained using this method in \S \ref{s:exp-vielbein}. We show explicit form of the metric expansion up to quartic order for $N=2, 3$ in \S \ref{s:metric-M^N}. We test our results using the direct method up to second order in Appendix \ref{a:verification}.

\subsection{The indirect method}
\label{s:indirect}

The indirect method is given as follows. The work of \cite{tubular} proves a general theorem which describes the tubular expansion of vielbein around $\cM$ sitting as a submanifold in an arbitrary ambient space $\cL$ to all orders. This result, as well as certain notations to be used below, are summarised in Appendix \ref{a:tubular}. We take this general result and specialise to our case where $\cL = \cM^N$ and the submanfiold under question is the diagonal one. At this stage the expansion coefficients, which are tensors of $\cM^N$ evaluated at the diagonal, are expressed in terms of FNC. Therefore the problem is to re-express all of them in terms of $\cM$-data. This can be done, thanks to their tensorial nature, by transforming them under $\hat z \to \bar z$ to express them in terms of tensors in DPC, which are directly known in terms of $\cM$-data. Notice that this requires a very limited amount of information about the coordinate transformation as the Jacobian matrix needs to be evaluated only at the submanifold. This is where the usefulness of the result in \cite{tubular} and the indirect method lies.

Below we will first compute this Jacobian matrix (restricted to the submanifold) following the general construction of \cite{FS} and then in \S \ref{s:exp-vielbein} we express all the quantities relevant to the expansion of vielbein in terms of $\cM$-data. Although this suffices for our practical goal, we perform certain further analysis for completeness of our overall understanding. This also facilitates the verification done in Appendix \ref{a:verification}. There are various steps of this analysis and the entire discussion has been kept in Appendix \ref{a:coord-construction}.

We now proceed to compute the relevant Jacobian matrix. Going back to eqs.(\ref{z-bar-gen}) and (\ref{zhat}), we note that without loss of generality we may take each one of $x$ and $x_p$ ($\forall p$) (labelling the points $Q$ and $l_p$ respectively) to be the general coordinate system $U$ as described in Appendix \ref{sa:RNC}. This implies that the metric in DPC has the following block-diagonal form\footnote{Following notations similar to that of Appendix \ref{a:tubular}, we use lower case symbols with a bar to denote tensors of $\cM^N$ in DPC. },
\bea
\bar g_{\bar a \bar b}(\zb) \to diag( {1\over N} G_{\alpha_1 \beta_1}(x_1), {1\over N} G_{\alpha_2 \beta_2}(x_2), \cdots , {1\over N} G_{\alpha_N \beta_N}(x_N))  ~,
\label{gbar}
\eea
where $G$ is the metric in $\cM$. The factor of $1/N$ on the RHS is due to the following reason. In an $N$-dimensional Cartesian system there exists an $N$-dependent scaling between the length scales along the axes and the diagonal. The above definition and the coordinate transformation to be discussed below will ensure that the induced metric on the diagonal submanifold be given by $G$. 
 
In order to define $\hat y^A$ in (\ref{zhat}), let us first denote the components of the tangent vector, 
\bea
(t_1, t_2, \cdots , t_N) \in T_{(x,x,\cdots , x)} \cM^N = T_x\cM \oplus \cdots \oplus T_x\cM   \quad (N\hbox{ additive factors})~,
\label{t-vector}
\eea 
in DPC by  $\bar \xi^{\bar a} = (\xi_1^{\al_1}, \xi_2^{\al_2}, \cdots , \xi_N^{\al_N})$. Then according to the relation between $l_p$ and $t_p$ as given by eq.(\ref{tp-lp}), we must have,
\bea
x_p^{\al_p} &=& \dlt^{\al_p}{}_{\al} x^{\al} + \Exp_x^{\al_p}(\xi_p) ~, \cr
&=& \dlt^{\al_p}{}_{\al} x^{\al} + \xi_p^{\al_p} + {\cal O}(\xi_p^2)~, 
\label{xp-xip}
\eea
where $\Exp_x: T_x\cM \to \cM$ \footnote{Since now onwards we will mostly work with coordinate description, by abuse of language, a point will usually be referred to by its coordinates in a given system. \label{pt-coord}} is the exponential map in $\cM$ as given in eq.(\ref{Exp}). According to the construction of \cite{FS}, $\hat y$ and $\bar \xi$ are linearly related,
\bea
\bar \xi^{\bar a} &=& \underline{K^{\bar a }{}_B} \hat y^B~,
\label{xibar-yhat}
\eea
where,
\bea
K^{\bar a}{}_b = \lt(\del \bar z^{\bar a}\over \del \hat z^b \rt) ~,
\eea
is the Jacobian matrix for the transformation $\hat z \to \bar z$. Notice that $\bar \xi$ satisfies the following constraint,
\bea
\sum_p \dlt^{\al}{}_{\al_p} \xi^{\al_p}_p = 0~,
\label{xi-transverse}
\eea
(which is the same equation in (\ref{sum-tp}) in component form) and therefore eq.(\ref{xibar-yhat}) is invertible.

Below we will first construct $J=K^{-1}$ by demanding that the metric in FNC at the location of the submanifold, as dictated by the results mentioned in Appendix \ref{a:tubular}, be given by,
\bea
\underline{\hat g_{a b}} &\to& diag(G_{\alpha \beta}(x), \eta_{AB})~, \quad \eta_{AB} = \eta_{\hat \al \hat \beta} \dlt_{\ha \hb} ~,
\label{ghat-submanifold}
\eea
where $\eta_{\al \beta}$, given the fact that $\cM$ is considered to be Riemannian, is simply given by the Kronecker delta. However, we will continue to use the symbol $\eta_{\al \beta}$ so that the expressions are generalizable to arbitrary signature. Before we go on to construct $J$, we explain the index notation adopted in the above equation. According to the index notation of eq.(\ref{zhat}), the transverse coordinates are denoted as 
$\hat y^{\hat \al_1}_1, \hat y^{\hat \al_2}_2 , \cdots $. We use the notation $\ha, \hb, \cdots$ to denote the subscript when it is arbitrary. But in that case we can make our notation less clumsy by removing the subscript from the tensor index (i.e. by writing $\hat y_{\ha}^{\hat \al}, \hat y_{\hb}^{\hat \beta}, \cdots$) but keeping in mind the association between Roman and Greek alphabets as $\ha \leftrightarrow \hat \al$, $\hb \leftrightarrow \hat \beta$ etc. At the same time the upper case Roman indices can be thought of being associated to pairs in the following way: $A \leftrightarrow (\ha, \hat \al)$, $B \leftrightarrow (\hb, \hat \beta)$.\footnote{
\label{indices}
Notice that the types of indices considered so far, namely $\alpha_p$, $\alpha$ and $\hat \alpha_{\ha}$ (or $\hat \al$) are all tangent space indices of the same manifold, i.e. $\cM$. Such indices are indistinguishable when they appear in a quantity that is intrinsic to $\cM$. However, they play different roles from the point of view of $\cM^N$.} Such an association is implied in the second equation of (\ref{ghat-submanifold}).

In order to construct $J$, we first look at the general way of decomposing a vector in $\cM^N$ into components that are tangential and normal to the diagonal submanifold. Given an arbitrary element in $T_{(x,x,\cdots , x)}\cM^N$ with components in DPC given by $\bar \eta^{\bar a} = (\eta_1^{\al_1}, \eta_2^{\al_2}, \cdots , \eta_N^{\al_N})$, one can define the tangential and normal parts as follows,
\bea
\bar \eta_{\parallel}^{\al} = R^{\al}{}_{\bar b} \bar \eta^{\bar b} ~, \quad \bar \eta_{\parallel}^A = 0~, 
\hbox{ and } \bar \eta_{\perp}^{\al} = 0 ~, \quad \bar \eta_{\perp}^A = R^A{}_{\bar b} \bar \eta^{\bar b} ~,
\eea
respectively, where $R = O \otimes \one_{\dim \cM}$ ($\one_{d}$ being the identity matrix of dimension $d$), i.e.
\bea
R^{\al}{}_{\beta_p} = O_{0p} \dlt^{\al}{}_{\beta_p} ~, \quad R^A{}_{\beta_p} = O_{\ha p} \dlt^{\hat \al}{}_{\beta_p}~, \quad \ha = 1, 2, \cdots (N-1)~, 
\label{R-def}
\eea 
and $O$ is an $N\times N$ orthogonal matrix such that,
\bea
O_{0p} &=& {1\over \sqrt{N}}~.
\label{O0p}
\eea
In fact, to preserve handedness of the coordinate system, we will always consider $O \in SO(N)$ in this work. 

The above definition of transversality and the existing direct product structure imply that,
\bea
\yh^A &=& {1\over \sqrt{N}} {\cal E}^A{}_B (x) R^B{}_{\bar b} \bar \xi^{\bar b} ~,
\label{yh-xibar}
\eea
where $\bar \xi^{\bar a}$ satisfy the condition (\ref{xi-transverse}) and the matrix ${\cal E}(x)$ has a block-diagonal form,
\bea
{\cal E}(x) = diag( {\cal E}_1(x), {\cal E}_2(x), \cdots , {\cal E}_{N-1}(x) )~,
\label{calE-def}
\eea
where the sub-matrix ${\cal E}_{\ha}(x)$ is unknown, to be determined below. Equations (\ref{yh-xibar}, \ref{xp-xip}) and (\ref{xibar-yhat}) imply,
\bea
\underline{J^{\al}{}_{\beta_p}} &=& {1\over \sqrt{N}} R^{\al}{}_{\beta_p} \quad   
\underline{J^A{}_{\beta_p}} = {1\over \sqrt{N}} {\cal E}^A{}_B(x) R^B{}_{\beta_p} ~,
\label{underline-J}
\\
\underline{K^{\al_p}{}_{\beta} } &=& \sqrt{N} (R^T)^{\al_p}{}_{\beta} ~, \quad 
\underline{K^{\al_p}{}_B} =  \sqrt{N} (R^T)^{\al_p}{}_C {\cal F}^C{}_B(x) ~, 
\eea
where,
\bea
{\cal F}(x) = diag ({\cal F}_1(x), {\cal F}_2(x), \cdots , {\cal F}_{N-1}(x) )~, 
\eea
such that,
\bea
{\cal F}_{\ha} (x) {\cal E}_{\ha}(x) &=& \one_{\dim \cM} ~.
\eea
Therefore to reach our final goal all we have to do is to find ${\cal E}_{\ha}(x)$ and ${\cal F}_{\ha}(x)$. This can be done by demanding that the coordinate transformation under consideration relate $\underline{\bar g_{\bar a \bar b}}$ as given in eq.(\ref{gbar}) and
$\underline{\hat g_{ab}}$ as given in eq.(\ref{ghat-submanifold}). This gives, upon using $O^TO=\one_N$,
\bea
{\cal E}_{\ha}^{\al}{}_{\beta} (x) &=& E^{(\al)}{}_{\beta}(x)~, \quad {\cal F}_{\ha}^{\al}{}_{\beta}(x) = E_{(\beta)}{}^{\al}(x) ~, \quad \forall \ha = 1, 2, \cdots , (N-1)~,
\label{calE}
\eea 
where $E^{(\al)}{}_{\beta}(x)$ is the vielbein of $\cM$ (see Appendix \ref{sa:RNC}). Notice that the use of indices in the above equations does not seem to be compatible with the rules mentioned below eq.(\ref{ghat-submanifold}). This is because those rules do not apply to ${\cal E}_{\ha}$ and ${\cal F}_{\ha}$ as these are quantities intrinsic to $\cM$ (see footnote \ref{indices}).

Our discussion so far enables one to relate any tensor in the two systems (FNC and DPC) at the submanifold. However, for a quantity which also caries an internal frame index, one has to find suitable basis for the frames as well in order to compare with the results summarized in Appendix \ref{a:tubular}. This is done simply by using the rotation matrix $R$. For example, given the vielbein components\footnote{The $N$-dependence of (\ref{ebar}) is obtained by requiring compatibility with (\ref{gbar}).},  
\bea
\bar e^{(\bar a)}{}_{\bar b} (\bar z) & \to & diag( {1\over \sqrt{N}} E^{(\al_1)}{}_{\beta_1}(x_1),  {1\over \sqrt{N}} E^{(\al_2)}{}_{\beta_2}(x_2), \cdots ) ~,
\label{ebar}
\eea
in DPC, we define the tangential and transverse components as,
\bea
\bar e_{\parallel}^{(\al)}{}_{\bar b} = R^{\al}{}_{\bar a} \bar e^{(\bar a)}{}_{\bar b}~, && \quad 
\bar e_{\perp}^{(A)}{}_{\bar b} = R^{A}{}_{\bar a} \bar e^{(\bar a)}{}_{\bar b}~.
\label{ebar-prl-perp}
\eea
Then the vielbein components in FNC are defined as follows,
\bea
\hat e^{(\al)}{}_b = \bar e_{\parallel}^{(\al)}{}_{\bar a} K^{\bar a}{}_b ~, && \quad 
\hat e^{(A)}{}_b = \bar e_{\perp}^{(A)}{}_{\bar a} K^{\bar a}{}_b ~,
\label{ehat-ebar-prl-perp}
\eea
It is the vielbein components of (\ref{ehat-ebar-prl-perp}) that we need to identify with the ones whose tubular expansion has been discussed in Appendix \ref{a:tubular}. The same prescription for defining tangential and transverse internal indices as given in eqs.(\ref{ebar-prl-perp}) is to be used for arbitrary tensors. For example, each term in the tubular expansion of $\hat e^{(a)}{}_b$ is of the form $\underline{\hat t^{(a)}{}_b}$, which should be written as (in matrix notation),
\bea
\underline{\hat{\hbox{t}}} &=& R \underline{\bar{\hbox{t}} K}~,
\label{that-tbar}
\eea
where $\bar{ \underline{\hbox{t}}}$ is the same tensor in DPC evaluated on the submanifold. Notice that the right hand side is entirely written in terms of $\cM$-data.

\subsection{Expansion coefficients for vielbein}
\label{s:exp-vielbein}

Given the above discussion, we can now compute all the quantities appearing in eqs.(\ref{FNC-ehat}) in terms of $\cM$-data using eq.(\ref{that-tbar}). However, there are a few points to be considered here.

The first one is to find the right $N$-dependence. To count the $N$-dependence systematically we introduce Weyl transformed tensors in DPC in the following manner. 
Just like for the metric (see eq.(\ref{gbar})), given any tensor $T$ in $\cM$, we construct a corresponding tensor $\bar t'$ in $\cM^N$ which is block-diagonal, such that the $p$-th block is given by,
\bea
t'_p(\bar z) = t'_p(x_p) = T(x_p) ~.
\label{t'-T}
\eea 
For example, for a tensor of rank, say $(2,1)$,
\bea
\bar t'^{\bar a \bar b}{}_{\bar c}(\bar z) &=& \lt\{ \begin{array}{ll}
t'_p{}^{\al_p \beta_p}{}_{\xi_p} (x_p) = T^{\al_p \beta_p}{}_{\xi_p} (x_p) 
& \hbox{for } \bar a = \al_p, \bar b = \beta_p, \bar c = \xi_p  \cr & \cr
0 & \hbox{otherwise }
\end{array} \rt\} ~, \forall p = 1, 2, \cdots, N ~. \cr &&
\label{example-tbar}
\eea
The above statements (\ref{t'-T}, \ref{example-tbar}) are, in fact, true not only for tensors, but for any quantity constructed out of vielbel and its derivatives. The primed tensors are related to the corresponding unprimed ones by a Weyl transformation. 
\bea
\bar{\hbox{t}} = N^{w\over 2} \bar {\hbox{t}}'~,
\label{weyl-weight}
\eea
where, $w$ is the Weyl-weight of the tensor. The above equation determines $N$-dependence of all the tensors. For example, $w=-2, 2, -1, 0$ for $\bar g_{\bar a\bar b}, \bar g^{\bar a \bar b}, \bar e^{(\bar a)}{}_{\bar b}$ and $\bar r^{\bar a}{}_{\bar b \bar c \bar d}$ respectively.

Our next concern is the following. The tubular expansion under consideration can be viewed as an expansion in powers of the vector $(t_1, t_2, \cdots , t_N)$ (see eq.(\ref{t-vector})). The expression for the expansion coefficients depends on the coordinate system chosen to describe this vector. For example, we could choose to use DPC, in which case the expansion parameter will be $\xi_p^{\al_p}$. Alternatively, we could also use FNC ($\hat y_{\ha}^{\hat \al}$) or any other coordinate system. The choice depends on the application. For example, if the (tubular) geometric structure of $\cM^N$ is appearing in a dynamical model in $\cM$, then it will be most suitable to expand in terms of $\xi_p$ as one is ultimately interested in a physical answer to be given completely in terms of $\cM$-data. On the other hand, recalling our discussion at the beginning of \S \ref{tub:M^N}, we may also view the tubular geometry under consideration as a natural Riemannian geometry on the bundle $T^{(N-1)}\cM$ where $\{ \hat y_{\ha}^{\hat \al} \}$ play the role of coordinates along fiber. From this point of view it will be natural to describe the geometry as an expansion in terms of $\{ \hat y_{\ha}^{\hat \al} \}$.

It turns out that the expressions look simpler if we use $\xi_p$ instead of $\hat y_{\ha}^{\hat \al}$. It will also turn out that this difference will not matter much when we extend the result to loop space in next section. Therefore below we choose to write the tubular expansion of vielbein in terms of $\xi_p$.

A typical term in this expansion is given by,
\bea
\underline{\hat t^{(a)}{}_{b D^1 \cdots D^n} } \hat y^{D^1} \cdots \hat y^{D^n} &=& R^a{}_{\bar a} 
\underline{\bar t^{(\bar a)}{}_{\bar b \bar d^1 \cdots \bar d^n} K^{\bar b}{}_b (K^{\bar d^1}{}_{D^1} J^{D^1}{}_{\bar e^1} ) \cdots (K^{\bar d^n}{}_{D^n} J^{D^n}{}_{\bar e^n} })  \bar \xi^{\bar e^1} \cdots \bar \xi^{\bar e^n}~, \cr
&=& N^{(w+1)\over 2} T^{(\hat \al)}{}_{(\hat \beta) \dlt^1 \cdots \dlt^n}(x) \sum_p O_{\ha p} O^T_{p \hb} \xi_p^{\dlt^1} \cdots \xi_p^{\dlt^n} ~, \ha, \hb = 0, 1,2, \cdots , \cr &&
\label{gen-coeff}
\eea
where we have used eq.(\ref{that-tbar}) and,
\bea
\underline{K^{\bar d}{}_A J^A{}_{\bar e}} \bar \xi^{\bar e} = \xi_p^{\dlt_p} ~.
\label{KJxi}
\eea
Notice that, to reduce clutter, in the second line we have specified the result for all values of the indices $a=(\al, A)$ and $b=(\beta, B)$ by allowing $\ha$ and $\hb$ to have the value $0$. According to our notation for indices, $\hat \al$ ($\hat \beta$) in the same equation should be replaced by $\al$ ($\beta$) whenever $\ha$ ($\hb$) possesses the value $0$.

Using the above results one can finally compute the expansion coefficients of vielbein. In addition to the last equation in (\ref{underline-e}), which remains the same, the final results are given by,
\bea
\hat e_0^{(\alpha)}{}_{\beta} &=& E^{(\alpha)}{}_{\beta}(x) ~, 
\label{ehat0-alpha} \\
\hat e_0^{(A)}{}_{\beta} &=& {1\over \sqrt{N}} \Omega_{\beta}{}^{(\hat \alpha)}{}_{\gamma}(x) 
\sum_p O_{\ha p} \xi^{\gamma}_p ~,
\label{ehat0-A} \\
\underline{\hat \pi^{(a)}{}_{(b)}}(\{s\}_n, \hat y) &=& \sum_p O_{\ha p} O^T_{p \hb} \Pi_x^{(\hat \alpha)}{}_{(\hat \beta)}(\{s\}_n, \xi_p) ~, \quad \ha, \hb = 0, 1, \cdots , (N-1)~. \cr &&
\label{pihat-lowerbar} 
\eea
The $\Pi$-matrix in the above equation is given by,
\bea
\Pi_x(\{s\}_n, \xi) &=& (\xi. \nabla)^{s_1} \cR(\xi;x) \cdots (\xi . \nabla)^{s_n} \cR(\xi;x)~, \cr
(\xi. \nabla)^s [\cR(\xi ;x)]^{(\alpha)}{}_{(\beta)} &=& \xi^{\alpha^1} \cdots \xi^{\alpha^s} \xi^{\gamma} \xi^{\dlt} \nabla_{\alpha^1} \cdots \nabla_{\alpha^s} R^{(\alpha)}{}_{\gamma \dlt (\beta)}(x) ~.
\label{Pihat}
\eea
Furthermore, 
\bea
\Omega_{\alpha}{}^{(\beta)}{}_{(\gamma)} &=& E^{(\beta)}{}_{\dlt} \nabla_{\alpha} E_{(\gamma)}{}^{\dlt} ~,
\label{spin-M}
\eea
$R^{\alpha}{}_{\beta \gamma \dlt}$ and $\nabla$ are spin connection, Riemann tensor and covariant derivative of $\cM$ respectively in the general coordinate system $U$ as described in Appendix \ref{sa:RNC}. The coordinate and non-coordinate indices are interchanged by the use of vielbein $E^{(\alpha)}{}_{\beta}$. For example, $\Omg_{\alpha}{}^{(\beta)}{}_{\gamma} = \Omg_{\alpha}{}^{(\beta)}{}_{(\dlt)} E^{(\dlt)}{}_{\gamma}$. Finally, notice that to reduce clutter we have packaged all the values of the indices $a$ and $b$ in eq.(\ref{pihat-lowerbar}) as was done in eq.(\ref{gen-coeff}).

\subsection{Some explicit results for metric-expansion}
\label{s:metric-M^N}

As we saw in the previous subsection, our method of computing tubular expansion of any tensor around $\Dlt \hookrightarrow \cM^N$ boils down to first writing down the expansion in the generic case of $\cM \hookrightarrow \cL$ and then specialize to $\cL = \cM^N$ and use the method as described in \S \ref{s:indirect} to express the results in terms of $\cM$-data. We follow the same procedure to arrive at explicit results for metric-expansion up to quartic order for $N=2, 3$. The necessary details of the computation are given in Appendix \ref{a:metric} where we also argue that $\Dlt \hookrightarrow \cM^N$ is totally geodesic for any $N$.

For $N=2$, we express the results in terms of $\hat y_1^{\hat \al} = \hat y^{\hat \al}$ (the index $\ha$ possesses only one value, i.e. $1$) and it can be interpreted as a natural Riemannian metric on tangent bundle $T\cM$. For $N=3$, to avoid complications we express the results in terms of $\bar \xi^{\bar a}$ which satisfies the constraint (\ref{xi-transverse}). 

\noindent
\underline{\bf $N=2$}

The $SO(2)$ matrix is uniquely fixed to be as given in eq.(\ref{SO2}). The final results are,
\bea
\hat g_{\alpha \beta} 
&=& G_{\alpha \beta} + (R_{\alpha (\hat \xi^1 \hat \xi^2) \beta} + \Omg_{\alpha}{}^{\eta}{}_{(\hat \xi^1)} \Omg_{\beta \eta (\hat \xi^2) } ) \yh^{\hat \xi^1 \hat \xi^2} 
+ \lt\{
{1\over 12} \nabla^{tot}_{(\hat \xi^1)}  \nabla_{(\hat \xi^2) } R_{\alpha (\hat \xi^3 \hat \xi^4) \beta}  
+ {1\over 3} R_{\alpha (\hat \xi^1 \hat \xi^2) \eta } R^{\eta}{}_{(\hat \xi^3 \hat \xi^4) \beta}  \rt. \cr
&& \lt.
+ {1\over 4 } ( \nabla^{tot}_{(\hat \xi^1)} R_{\eta (\hat \xi^2 \hat \xi^3) \beta} \Omg_{\alpha}{}^{\eta}{}_{(\hat \xi^4)}  
+ \alpha \leftrightarrow \beta ) 
+ {1\over 3} R_{\eta (\hat \xi^1 \hat \xi^2) \dlt } \Omg_{\alpha}{}^{\eta}{}_{(\hat \xi^3)} \Omg_{\beta}{}^{\dlt}{}_{(\hat \xi^4)}  \rt\} \yh^{\hat \xi^1 \cdots \hat \xi^4} + O(\yh^5) ~, 
\label{g-al-beta-2} \\
\hat g_{\alpha \hat \beta } &=& \Omg_{\alpha}{}_{(\hat \beta \hat \xi)} \yh^{\hat \xi} 
+ \lt\{ {1\over 4} \nabla_{(\hat \xi^1)} R_{\alpha (\hat \xi^2 \hat \xi^3 \hat \beta) } 
+ {1\over 3} \Omg_{\alpha}{}^{\eta}{}_{(\hat \xi^1)} R_{\eta (\hat \xi^2 \hat \xi^3 \hat \beta) } \rt\} \yh^{\hat \xi^1 \cdots \hat \xi^3}  + O(\yh^5) ~, 
\label{g-al-B-2} \\
\hat g_{\hat \al \hat \beta}
&=& \eta_{\hat \al \hat \beta } + {1\over 3} R_{(\hat \al \hat \xi^1 \hat \xi^2 \hat \beta) } \yh^{\hat \xi^1} 
\yh^{\hat \xi^2} + ( {1\over 20} \nabla_{(\hat \xi^1)} \nabla_{(\hat \xi^2)} R_{( \hat \al \hat \xi^3 \hat \xi^4 \hat \beta) } + {2\over 45} R_{(\hat \al \hat \xi^1 \hat \xi^2) \eta } R^{\eta}{}_{(\hat \xi^3 \hat \xi^4 \hat \beta) } ) \yh^{\hat \xi^1 \cdots \hat \xi^4}  \cr
&& + O(\yh^5) ~,
\label{g-A-B-2}
\eea
where we have used the notation: $\hat y^{\hat \xi^1 \cdots \hat \xi^n} = \hat y^{\hat \xi^1} \cdots \hat y^{\hat \xi^n}$.
The geometric quantities appearing on the RHS, namely $G$, $\Omega$, $R$ and its covariant derivatives are all evaluated at 
$x\in \Dlt \cong \cM$. Also note that according to our notations as explained below eq.(\ref{spin-M}),
\bea
\nabla_{(\xi^1)} R_{\alpha (\xi^2 \xi^3) \beta} 
&=& E_{(\xi^1)}{}^{\eta^1} \nabla^{tot}_{\eta^1} (E_{(\xi^2)}{}^{\eta^2} E_{(\xi^3)}{}^{\eta^3}  R_{\alpha \eta^2 \eta^3 \beta} ) ~,
\eea
where $\nabla^{tot}$ is the total covariant derivative which annihilates vielbein.

\noindent
\underline{\bf $N=3$} 

The $SO(3)$ matrix is taken to be as given in eq.(\ref{SO3}). This is of course not a unique choice. The final results are given by,
\bea
\hat g_{\al \beta} 
&=& G_{\alpha \beta} + {1\over 3} \lt[ (  R_{\alpha \gamma^1 \gamma^2 \beta} 
+ \Omg_{\alpha}{}^{\eta}{}_{\gamma^1} \Omg_{\beta \eta \gamma^2} ) \sum_p \xi_p^{\gamma^1 \gamma^2} \rt. \cr
&& + ( {1\over 3} \nabla^{tot}_{\gamma^1} R_{\alpha \gamma^2 \gamma^3 \beta} 
+ {2\over 3} R_{\alpha \gamma^1 \gamma^2 \eta} \Omg_{\beta}{}^{\eta}{}_{\gamma^3}  
+ {2\over 3} R_{\beta \gamma^1 \gamma^2 \eta} \Omg_{\alpha}{}^{\eta}{}_{\gamma^3}  ) \sum_p \xi_p^{\gamma^1 \cdots \gamma^3} \cr
&& + \lt\{ {1\over 12} \nabla^{tot}_{\gamma^1 \gamma^2} R_{\alpha \gamma^3 \gamma^4 \beta} 
+ {1\over 3 } R_{\alpha \gamma^1 \gamma^2 \eta} R^{\eta}{}_{\gamma^3 \gamma^4 \beta} 
+ {1\over 4 } ( \nabla_{\gamma^1} R_{ \al \gamma^2 \gamma^3 \eta } \Omg_{\beta }{}^{\eta}{}_{\gamma^4 } 
+ \alpha \leftrightarrow \beta ) \rt. \cr
&& \lt. \lt. 
+ {1\over 3} R_{\eta \gamma^3 \gamma^4 \dlt} \Omg_{\alpha}{}^{\eta}{}_{\gamma^1} \Omg_{\beta}{}^{\dlt}{}_{\gamma^2}  \rt\} 
\sum_p \xi_p^{\gamma^1 \cdots \gamma^4} + O(\xi^5) \rt] ~.   \cr
&=& G_{\alpha \beta} + {1\over 3} (  R_{\alpha \gamma^1 \gamma^2 \beta} 
+ \Omg_{\alpha}{}^{\eta}{}_{\gamma^1} \Omg_{\beta \eta \gamma^2} ) \sum_p \xi_p^{\gamma^1 \gamma^2}  \cr
&& + ( {1\over 9} \nabla^{tot}_{\gamma^1} R_{\alpha \gamma^2 \gamma^3 \beta} 
+ {2 \over 9}  R_{\alpha \gamma^1 \gamma^2 \eta} \Omg_{\beta}{}^{\eta}{}_{\gamma^3}  
+ {2 \over 9}  R_{\beta \gamma^1 \gamma^2 \eta} \Omg_{\alpha}{}^{\eta}{}_{\gamma^3}  ) \sum_p \xi_p^{\gamma^1 \cdots \gamma^3} \cr
&& + \lt\{ {1\over 36 } \nabla^{tot}_{\gamma^1} \nabla^{tot}_{ \gamma^2} R_{\alpha \gamma^3 \gamma^4 \beta} 
+ {1\over 12 } R_{\alpha \gamma^1 \gamma^2 \eta} R^{\eta}{}_{\gamma^3 \gamma^4 \beta} 
+ {1\over 12 } ( \nabla^{tot}_{\gamma^1} R_{ \al \gamma^2 \gamma^3 \eta } \Omg_{\beta }{}^{\eta}{}_{\gamma^4 } 
+ \alpha \leftrightarrow \beta ) \rt. \cr
&& \lt.  
+ {1\over 9 } R_{\eta \gamma^3 \gamma^4 \dlt} \Omg_{\alpha}{}^{\eta}{}_{\gamma^1} \Omg_{\beta}{}^{\dlt}{}_{\gamma^2}  \rt\} 
\sum_p \xi_p^{\gamma^1 \cdots \gamma^4} + O(\xi^5)  ~,  
\label{g-al-beta-3}
\\
&& \cr
\hat g_{\al \hat \beta_1} &=& {1\over \sqrt{6}} \lt[ \Omg_{\alpha}{}_{(\hat \beta_1) \gamma } (-\xi_1^{\gamma} + \xi_2^{\gamma}) 
+ {2\over 3} R_{\alpha \gamma^1 \gamma^2 (\hat \beta_1)} (- \xi_1^{\gamma^1 \gamma^2} + \xi_2^{\gamma^1 \gamma^2}) 
\rt. \cr
&&
+ ( {1\over 4} \nabla^{tot}_{\gamma^1} R_{\alpha \gamma^2 \gamma^3 (\hat \beta_1) } 
+ {1 \over 3} \Omg_{\alpha}{}^{\eta}{}_{\gamma^1} R_{\eta \gamma^2 \gamma^3 (\hat \beta_1) }  ) 
(-\xi_1^{\gamma^1 \cdots \gamma^3} + \xi_2^{\gamma^1 \cdots \gamma^3} ) \cr
&& 
+ ( {1\over 15} \nabla^{tot}_{\gamma^1} \nabla^{tot}_{\gamma^2} R_{\alpha \gamma^3 \gamma^4 (\hat \beta_1) } 
+ {2\over 15} R_{\alpha \gamma^1 \gamma^2 \eta} R^{\eta}{}_{\gamma^3 \gamma^4 (\hat \beta_1) } 
+ {1\over 6} \Omg_{\alpha}{}^{\eta}{}_{\gamma^1} \nabla^{tot}_{\gamma^2} R_{\eta \gamma^3 \gamma^4 (\hat \beta_1) } ) 
(- \xi_1^{\gamma^1 \cdots \gamma^4} + \xi_2^{\gamma^1 \cdots \gamma^4}) \cr
&& \lt. + O(\yh^5) \rt] ~, 
\label{g-al-betahat1-3} \\
\hat g_{\al \hat \beta_2} &=&  {1\over 3\sqrt{2}} \lt[ \Omg_{\alpha}{}_{(\hat \beta_2) \gamma } 
(-\xi_1^{\gamma} - \xi_2^{\gamma} + 2 \xi_3^{\gamma}) 
+ {2\over 3} R_{\alpha \gamma^1 \gamma^2 (\hat \beta_2 )} (- \xi_1^{\gamma^1 \gamma^2} 
- \xi_2^{\gamma^1 \gamma^2} + 2 \xi_3^{\gamma^1 \gamma^2} ) \rt. \cr
&& + ( {1\over 4} \nabla^{tot}_{\gamma^1} R_{\alpha \gamma^2 \gamma^3 (\hat \beta_2 ) } 
+ {1 \over 3} \Omg_{\alpha}{}^{\eta}{}_{\gamma^1} R_{\eta \gamma^2 \gamma^3 (\hat \beta_2) } ) 
(-\xi_1^{\gamma^1 \cdots \gamma^3} - \xi_2^{\gamma^1 \cdots \gamma^3} + 2 \xi_3^{\gamma^1 \cdots \gamma^3}) \cr
&& + ( {1\over 15} \nabla^{tot}_{\gamma^1} \nabla^{tot}_{\gamma^2} R_{\alpha \gamma^3 \gamma^4 (\hat \beta_2 ) } 
+ {2\over 15} R_{\alpha \gamma^1 \gamma^2 \eta} R^{\eta}{}_{\gamma^3 \gamma^4 (\hat \beta_2 ) }  \cr 
&& \lt. 
+ {1\over 6 } \Omg_{\alpha}{}^{\eta}{}_{\gamma^1} \nabla^{tot}_{\gamma^2} R_{\eta \gamma^3 \gamma^4 (\hat \beta_2 ) } ) 
(- \xi_1^{\gamma^1 \cdots \gamma^4} - \xi_2^{\gamma^1 \cdots \gamma^4} + 2 \xi_3^{\gamma^1 \cdots \gamma^4}) + O(\yh^5) \rt] ~, 
\label{g-al-betahat2-3}
\\
&& \cr
\hat g_{\hat \al_1 \hat \beta_1}
&=& \eta_{\hat \al_1 \hat \beta_1} + {1\over 2} \lt[ {1\over 3} R_{(\hat \al_1) \gamma^1 \gamma^2 (\hat \beta_1) } 
(\xi_1^{\gamma^1 \gamma^2} + \xi_2^{\gamma^1 \gamma^2}) 
+ {1\over 6} \nabla^{tot}_{\gamma^1} R_{(\hat \al_1) \gamma^2 \gamma^3 (\hat \beta_1)} 
(\xi_1^{\gamma^1 \cdots \gamma^3} + \xi_2^{\gamma^1 \cdots \gamma^3}) \rt. \cr
&& \lt. + ( {1\over 20} \nabla^{tot}_{\gamma^1} \nabla^{tot}_{\gamma^2} R_{(\hat \al_1) \gamma^3 \gamma^4 (\hat \beta_1) }  + {2\over 45} R_{(\hat \al_1) \gamma^1 \gamma^2 \eta } R^{\eta}{}_{\gamma^3 \gamma^4 (\hat \beta_1)}  )
(\xi_1^{\gamma^1 \cdots \gamma^4} + \xi_2^{\gamma^1 \cdots \gamma^4}) + O(\yh^5) \rt] ~, \cr && 
\label{g-alhat1-betahat1-3}
\\
\hat g_{\hat \al_1 \hat \beta_2}
&=& {1\over 2 \sqrt{3} } \lt[ {1\over 3} R_{(\hat \al_1) \gamma^1 \gamma^2 (\hat \beta_2) } 
(\xi_1^{\gamma^1 \gamma^2} - \xi_2^{\gamma^1 \gamma^2}) 
+ {1\over 6} \nabla^{tot}_{\gamma^1} R_{(\hat \al_1) \gamma^2 \gamma^3 (\hat \beta_2)} 
(\xi_1^{\gamma^1 \cdots \gamma^3} - \xi_2^{\gamma^1 \cdots \gamma^3}) \rt. \cr
&& \lt. + ( {1\over 20} \nabla^{tot}_{\gamma^1} \nabla^{tot}_{\gamma^2} R_{(\hat \al_1) \gamma^3 \gamma^4 (\hat \beta_2) }  
+ {2\over 45} R_{(\hat \al_1) \gamma^1 \gamma^2 \eta } R^{\eta}{}_{\gamma^3 \gamma^4 (\hat \beta_2 )}  ) (\xi_1^{\gamma^1 \cdots \gamma^4} - \xi_2^{\gamma^1 \cdots \gamma^4}) + O(\yh^5) \rt] ~, \cr && 
\label{g-alhat1-betahat2-3}
\\
\hat g_{\hat \al_2 \hat \beta_2}
&=& \eta_{\hat \al_2 \hat \beta_2} 
+ {1\over 6 } \lt[ {1\over 3} R_{(\hat \al_2) \gamma^1 \gamma^2 (\hat \beta_2) } 
(\xi_1^{\gamma^1 \gamma^2} + \xi_2^{\gamma^1 \gamma^2} + 4  \xi_3^{\gamma^1 \gamma^2} )  \rt. \cr
&& + {1\over 6} \nabla^{tot}_{\gamma^1} R_{(\hat \al_1) \gamma^2 \gamma^3 (\hat \beta_1)} 
(\xi_1^{\gamma^1 \cdots \gamma^3} + \xi_2^{\gamma^1 \cdots \gamma^3} + 4 \xi_3^{\gamma^1 \cdots \gamma^3} )   \cr
&& + ( {1\over 20} \nabla^{tot}_{\gamma^1} \nabla^{tot}_{\gamma^2} R_{(\hat \al_1) \gamma^3 \gamma^4 (\hat \beta_1) }  + {2\over 45} R_{(\hat \al_1) \gamma^1 \gamma^2 \eta } R^{\eta}{}_{\gamma^3 \gamma^4 (\hat \beta_1)}  ) (\xi_1^{\gamma^1 \cdots \gamma^4} + \xi_2^{\gamma^1 \cdots \gamma^4} + 4 \xi_3^{\gamma^1 \cdots \gamma^4} ) \cr
&& \lt. + O(\yh^5) \rt] ~. 
\label{g-alhat2-betahat2-3}
\eea

\section{Tubular geometry in $\cL\cM$}
\label{tub:LM}

We would now like to construct the tubular geometry of $\cL\cM$ near the submanifold of vanishing loops by performing a suitable large-$N$ limit of the construction described in the previous section. Note, however, that metric in $\cL \cM$ is well-known. We first show in \S \ref{s:LM-DPC} that this well-known form is nothing but the large-$N$ limit of the metric of $\cL^{(N)}\cM$ in DPC, i.e. the analogue of eq.(\ref{gbar}). Our goal here is to find analogues of eqs.(\ref{ehat0-alpha}-\ref{pihat-lowerbar}) and (\ref{g-al-beta-3}-\ref{g-alhat2-betahat2-3}) which will be done in \S \ref{s:FNC-LM} and \S \ref{s:metric-LM} respectively.

\subsection{Geometry of $\cL\cM$ and DPC}
\label{s:LM-DPC}

We discussed in \S \ref{s:exp-vielbein} how given a tensor in $\cM$, one can construct a corresponding tensor in $\cM^N$ in DPC. All the geometric quantities of $\cM^N$, which are constructed out of vielbein and its derivatives, are of this type, as required by the discrete isometries (to be discussed in \S \ref{s:isometry}). All computations involving such tensors are expressible in terms of $\cM$-data. Let $\bar v^{\bar a}(\bar z)$ and $V^{\al}(x)$ be the components of the corresponding tangent vector fields in $\cM^N$ and $\cM$ respectively. We relate them in the following way (see \S \ref{s:exp-vielbein} for notation),
\bea
v_p^{\al_p}(x_p) &=& V^{\al_p}(x_p)~, 
\eea
such that the lengths match on the diagonal ($x_p=x~, \forall p$),
\bea
|\bar v (\bar z)|^2|_{x_p=x} &\equiv& \bar v^{\bar a}(\bar z) \bar v^{\bar b}(\bar z) \bar g_{\bar a \bar b}(\bar z) |_{x_p=x} = |V(x)|^2 ~.
\label{v-mod-squared}
\eea

The above discussion is also valid for $\cL^{(N)}\cM$ as given in (\ref{cut-off}). Restricting ourselves to this space, we now proceed to consider the large-$N$ limit. To this end, we introduce,
\bea
\s &=& {2\pi \over m} (p-1)~, \quad m = 2N+1~, 
\label{sigma}
\eea
which becomes, at large $N$, a continuous parameter with range from $0$ to $2\pi$ as $p$ goes from $1$ to $m$. We will identify this as the internal parameter of the loop. Moreover, at large $N$, we will restrict the values of $\{x_p\}$ to be such that the loop is smooth. This implies that the DPC given in eq.(\ref{z-bar-gen}) takes the following form at large $N$,
\bea
\bar z^{\bar a} &\to & x^{\al}(\s)~,
\label{zbar-xsigma}
\eea
where $x(\s) \in C^{\infty}(S^1, \cM)$. Notice that the coordinate index $\al$ in eq.(\ref{z-bar-gen}) carried a discrete subindex corresponding to different copies of $\cM$. This has now become a continuous parameter and we have removed it in the above equation for simplicity, with the (usual) understanding that the value of $\al$ is chosen independently for different values of $\s$. It is now clear how our notations for DPC must be transformed into the usual {\it loopy} notations in the large-$N$ limit. For example, for the tensor in (\ref{example-tbar}) we must have,
\bea
t'_p{}^{\al_p \beta_p}{}_{\xi_p} (x_p) &\to&  T^{\al \beta}{}_{\xi}(x(\s))~.
\eea
Therefore the norm in eq.(\ref{v-mod-squared}) takes the following form,
\bea
|\bar v(\bar z)|^2 &\to& \oint {d\s \over 2\pi} V^{\al}(x(\s)) V^{\beta}(x(\s)) G_{\al \beta}(x(\s)) ~,
\label{LM-metric-usual}
\eea
where we have used the following large-$N$ property,
\bea
{1\over m} \sum_{p=1}^m \cdots &\to&  \oint {d\s \over 2\pi} \cdots ~. 
\label{sum-int}
\eea
Equation (\ref{LM-metric-usual}) is the standard way of specifying the metric on $\cL\cM$ and we have shown here how this description is related to a large-$N$ limit in DPC. 

\subsection{FNC and tubular geometry}
\label{s:FNC-LM}

Here we will implement the large-$N$ limit in the tubular construction as discussed in \S \ref{tub:M^N} applied to $\cL^{(N)}\cM$. While the general approach of \S \ref{tub:M^N} remains the same, we will incorporate certain important modifications. 
  
Recall that to identify FNC, it was very crucial to first separate out directions which are tangential and orthogonal to the submanifold and then to scale the orthogonal coordinates in such a way that the transverse part of the metric is flat at the leading order everywhere on the submanifold - see eq.(\ref{ghat-submanifold}). This separation was done by using an orthogonal matrix which made the transverse coordinates real. However, if we allow the transverse coordinates to be complex, then the same can also be achieved through a specific unitary matrix corresponding to a discrete Fourier transform. In the large-$N$ limit such coordinates correspond to non-zero left and right moving momentum modes on the loop in $T\cM$-description as explained below eq.(\ref{oint-t}). It is this system that we are going to use to describe the tubular geometry in $\cL\cM$.

Below we list the steps to be followed in order to translate tubular expressions in $\cL^{(N)}\cM$  to the corresponding ones in $\cL \cM$.
\begin{enumerate} 
\item {\bf Complex FNC:} Replace orthogonal matrix $O$ by a unitary matrix $U$ (to be given below) and $O^T$ by $U^{\dagger}$.\footnote{The matrix $O^T$ appeared in various expressions because of the involvement of $\underline{K} = \underline{J}^{-1}$, which now contains $U^{\dagger}$. } The resultant FNC is complex and we rename the transverse indices in the following way,
\bea
A \to (\hat \alpha, \ha)~, \quad \ha = - N, - N + 1, \cdots , (\neq 0), \cdots N-1, N ~.
\eea  

\item {\bf Discrete Fourier Transform (DFT) on $(\cM^{2N+1})_C$:} Use the following expressions for the unitary matrix elements,
\bea
U_{0p} &=& {1\over \sqrt{m}} ~, \quad U_{\ha p} = {1\over \sqrt{m}} e^{-{2\pi i\over m} (p-1) \ha}~.
\label{U-def}
\eea

\item
{\bf Large $N$/continuum limit:} After introducing the parameter in eq.(\ref{sigma}), we take the large-$N$ limit. 
\begin{itemize}
\item
In this limit the unitarity conditions are preserved in the following manner,
\bea
\sum_p U_{\ha p} (U^{\dagger})_{p\ha'} &\to & \oint {d\s \over 2\pi} e^{-i(\ha -\ha')\s} = \dlt_{\ha, \ha'}~, \label{sum-p} \\
\sum_{\ha =- N}^{N} (U^{\dagger})_{p\ha} U_{\ha p'} = {1\over m} \sum_{\ha = -N}^N e^{{2\pi i\over m}\ha (p-p')} &\to & {1\over 2\pi} \sum_{\ha \in \ZZ} e^{i\ha (\s-\s')} = \dlt(\s-\s')~. \cr &&
\label{sum-a}
\eea
While (\ref{sum-p}) directly follows from (\ref{sum-int}), the limit in (\ref{sum-a}) is true in the following sense,
\bea
\sum_p \lt( {1\over m} \sum_{\ha = -N}^N e^{{2\pi i\over m}\ha (p-p')} \rt) = 1 =  \sum_{\ha \in \ZZ} \oint {d\s \over 2\pi} e^{i\ha (\s-\s')}  ~. 
\eea

\item
Finally, we follow the general prescription of transiting from the discrete DPC-notations to the usual loopy-notations as discussed in \S \ref{s:LM-DPC}. For example, for $\xi_p$ defined near eq.(\ref{xi-transverse}) we have,
\bea
\xi_p^{\alpha_p} \to \xi^{\alpha}(\s) ~, \quad \hbox{such that} \quad \oint d\s \xi^{\alpha}(\s) = 0~.
\eea
Therefore following eq.(\ref{yh-xibar}), the FNC should read,
\bea
\hat y^A &=& {1\over \sqrt{m}} E^{(\hat \al)}{}_{\beta}(x) \sum_p U_{\ha p} \xi_p^{\beta} \to  \oint {d\s \over 2\pi} e^{-i\ha \s} \hat Y^{\hat \al}(\s)~,
\label{yhat-Yhat}
\eea
where $\hat Y^{\hat \al}(\s) = E^{(\hat \al)}{}_{\beta}(x) \xi^{\beta}(\s)$. Therefore the latter is the description of the loop in RNC (see eq.(\ref{Yh-Y'}, \ref{Y'-xi})) centred at $x \in \cM$, the latter being the CM of the loop.

\end{itemize}

\end{enumerate}

Following the above steps one can re-work-out the expressions analogous to those in eqs.(\ref{ehat0-A}, \ref{pihat-lowerbar}). The final results are as follows,
\bea
\hat e_0^{(A)}{}_{\beta} 
&=& \Omega_{\beta}{}^{(\hat \alpha)}{}_{\dlt}(x) \oint {d\s \over 2\pi} e^{-i\ha \s} \xi^{\dlt}(\s)~, \cr
\underline{\hat \pi^{(a)}{}_{(b)}}(\{s\}_n, \hat y) &=& \oint {d\s \over 2\pi}  
e^{-i(\ha -\hb)\s} \Pi_x^{(\hat \alpha)}{}_{(\hat \beta)} (\{s\}_n,\xi (\s))~, \quad \ha, \hb \in \ZZ ~, 
\eea
where just like in (\ref{pihat-lowerbar}), to reduce clutter we have combined four equations into one by allowing the indices $\ha$ and $\hb$ to take the value $0$. 

Similar rules were suggested relating geometric quantities in $\cL\cM$ and the corresponding ones in $\cM$ in general coordinates in earlier work \cite{dwv}. What we suggest here is that such rules better be defined for tubular expressions. This implies that in order to express a geometric quantity of $\cL \cM$ written in general coordinates in terms of $\cM$-data, one may first write it in the form of tubular expansion and then evaluate each term in the expansion in terms of $\cM$-data following the above procedure.

\subsection{Metric-expansion up to quartic order}
\label{s:metric-LM}

In order to perform tubular expansion of any tensor in $\cL\cM$, one uses a similar method as described at the beginning of \S \ref{s:metric-M^N}. The difference is that now one specialises to $\cL=\cL^{(N)}\cM$, replaces $O$ by $U$ etc and finally performs the continuum limit. The relevant details for the metric-expansion are given in Appendix \ref{a:metric}. The final results are given below,
\bea
\hat g_{\alpha \beta} 
&=& G_{\alpha \beta} + ( R_{\alpha (\hat \xi^1 \hat \xi^2) \beta} + \Omg_{\alpha}{}^{\eta}{}_{(\hat \xi^1)} \Omg_{\beta \eta (\hat \xi^2)}  ) \oint {d\s \over 2\pi} \Yh^{\hat \xi^1 \hat \xi^2 }(\s) \cr
&& +( {1\over 3} \nabla^{tot}_{(\hat \xi^1)} R_{\alpha (\hat \xi^2 \hat \xi^3) \beta} 
+ {2\over 3} R_{\alpha (\hat \xi^1 \hat \xi^2) \eta } \Omg_{\beta}{}^{\eta}{}_{(\hat \xi^3) }  
+ {2\over 3} R_{\beta (\hat \xi^1 \hat \xi^2) \eta} \Omg_{\alpha}{}^{\eta}{}_{(\hat \xi^3)}  ) 
\oint {d\s \over 2\pi} \Yh^{\hat \xi^1 \cdots \hat \xi^3}(\s)  \cr
&& + \lt\{ 
{1\over 12} \nabla^{tot}_{(\hat \xi^1)}  \nabla^{tot}_{(\hat \xi^2)} R_{\alpha (\hat \xi^3 \hat \xi^4) \beta} 
+ {1\over 3} R_{\alpha (\hat \xi^1 \hat \xi^2) \eta} R^{\eta}{}_{(\hat \xi^3 \hat \xi^4) \beta}  \rt. \cr
&& \lt.
+ {1\over 4 } ( \nabla^{tot}_{(\hat \xi^1)} R_{\alpha (\hat \xi^2 \hat \xi^3) \eta} \Omg_{\beta}{}^{\eta}{}_{(\hat \xi^4)} 
+ \alpha \leftrightarrow \beta )
+ {1\over 3} R_{\eta (\hat \xi^3 \hat \xi^4) \zeta} \Omg_{\alpha}{}^{\eta}{}_{(\hat \xi^1)} \Omg_{\beta}{}^{\zeta}{}_{(\hat \xi^2)}  \rt\} 
\oint {d\s \over 2\pi} \Yh^{\hat \xi^1 \cdots \hat \xi^4} (\s) + O(\hat Y^5) ~, \cr &&
\label{g-al-bet-LM}
\\
g_{\alpha B} 
&=& \Omg_{\alpha}{}_{(\hat \beta \hat \xi)} \oint {d\s \over 2\pi} e^{i \hb \s} \Yh^{\hat \xi}(\s)  
+ {2\over 3} R_{\alpha (\hat \xi^1 \hat \xi^2 \hat \beta)} \oint {d\s \over 2\pi} e^{i\hb \s} \Yh^{\hat \xi^1 \hat \xi^2}(\s) \cr
&& + ( {1\over 4} \nabla^{tot}_{(\hat \xi^1)} R_{\alpha (\hat \xi^2 \hat \xi^3 \hat \beta)} 
+ {1\over 3} \Omg_{\alpha}{}^{\eta}{}_{(\hat \xi^1)} R_{\eta (\hat \xi^2 \hat \xi^3 \hat \beta) } ) \oint {d\s \over 2\pi} e^{i\hb \s} \Yh^{\hat \xi^1 \cdots \hat \xi^3}(\s) \cr
&& + ( {1\over 15} \nabla^{tot}_{(\hat \xi^1)} \nabla^{tot}_{(\hat \xi^2)} R_{\alpha (\hat \xi^3 \hat \xi^4 \hat \beta)} 
+ {2\over 15} R_{\alpha (\hat \xi^1 \hat \xi^2) \eta} R^{\eta}{}_{(\hat \xi^3 \hat \xi^4 \hat \beta) }  \cr
&& + {1\over 6 } \Omg_{\alpha}{}^{\eta}{}_{(\hat \xi^1)} \nabla^{tot}_{(\hat \xi^2)} R_{\eta (\hat \xi^3 \hat \xi^4 \hat \beta) } ) \oint {d\s \over 2\pi} e^{i\hb \s} \Yh^{\hat \xi^1 \cdots \hat \xi^4}(\s)  + O(\hat Y^5)  ~,  
\label{g-al-B-LM} \\
g_{AB}
&=& \eta_{AB} + {1\over 3} R_{(\hat \alpha \hat \xi^1 \hat \xi^2 \hat \beta)} \oint {d\s \over 2\pi} e^{i(\ha + \hb)\s} \Yh^{\hat \xi^1 \hat \xi^2}(\s) \cr
&& + {1\over 6} \nabla^{tot}_{(\hat \xi^1)} R_{(\hat \alpha \hat \xi^2 \hat \xi^3 \hat \beta)} \oint {d\s \over 2\pi} e^{i(\ha + \hb) \s} \Yh^{\hat \xi^1 \cdots \hat \xi^3}(\s)   \cr
&& + ( {1\over 20} \nabla^{tot}_{(\hat \xi^1)} \nabla^{tot}_{(\hat \xi^2)} R_{(\hat \alpha \hat \xi^3 \hat \xi^4 \hat \beta)} + {2\over 45} R_{(\hat \alpha \hat \xi^1 \hat \xi^2) \eta} R^{\eta}{}_{(\hat \xi^3 \hat \xi^4 \hat \beta)} ) \oint {d\s\over 2\pi} e^{i(\ha + \hb)\s} \Yh^{\hat \xi^1 \cdots \hat \xi^4}(\s)  \cr
&& + O(\hat Y^5) ~.
\label{g-A-B-LM}
\eea
where we have used the notation: $\hat Y^{\hat \al^1 \hat \al^2 \cdots}(\s) \equiv \hat Y^{\hat \al^1}(\s) \hat Y^{\hat \al^2}(\s) \cdots$ and
\bea
\eta_{AB} &=& \eta_{\hat \alpha \hat \beta} \dlt_{\ha + \hb, 0}~.
\eea
Notice that the above expression is different from the one in (\ref{ghat-submanifold}). This is simply because of the complex coordinates chosen here.

\subsection{Isometry}
\label{s:isometry}

Our discussion in this section so far shows how, given the tubular geometry around $\Dlt \hookrightarrow {\cal L}^{(N)}\cM$, a specific large-$N$ limit can be defined. As mentioned earlier, in order for it be the right tubular geometry of $\cL \cM$, it must satisfy the requirement of isometry. Here we will first discuss in \S \ref{ss:discrete} the discrete isometry of ${\cal L}^{(N)}\cM$ and show how the reparametrization isometry arises in the large-$N$ limit. Then in \S \ref{ss:reparametrization} we will show that our large-$N$ geometry indeed satisfies the required Killing equation (in vielbein form) to all orders in tubular expansion.

\subsubsection{Discrete isometry and continuum limit}
\label{ss:discrete} 

The discrete isometries of $\cM^N$ that are independent of $\cM$ are the ones that 
permute factors of $\cM$ in $\cM^N$. 
The transformation is given by,
\bea
\bar z^{\bar a} \to \bar z'^{\bar a'} = (\cS_N^{-1})^{\bar a'}{}_{\bar b} \bar z^{\bar b}
\label{perm-transf}
\eea
where $\cS_N$ is the following matrix,
\bea
\cS_N = S_N \otimes \one_d~, 
\label{cS}
\eea
$S_N$ being an $N \times N$ permutation matrix, i.e.,
\bea
(S_N)_{pq} &=& \dlt_{\omg(p), q} = \dlt_{p, \omg^{-1}(q)} ~, 
\eea
where $\omg : \{ 1, 2, \cdots , N\} \to \{ 1, 2, \cdots , N\} $ is a bijection. Using eq.(\ref{ebar}), it is straightforward to show,
\bea
\bar e'(z') &=& \bar e(z')~,
\eea
as required.

For an ordered product $(\cM^m)_C$, only cyclic permutations remain as isometries.
This is given by replacing $S_N$ above by the following,
\bea
(C_N)_{pq} &=& \dlt_{\psi(p), q}~,
\label{C-N}
\eea
where,
\bea
\psi(p) &=& p + n ~ \hbox{mod } (m+1)~, \quad \forall n \in \{1, 2, \cdots , m \}~.  
\label{psi-p}
\eea

We now look at the large-$N$ limit. The cyclic permutation becomes constant translation in terms of the parameter $\s$, 
\bea
\s \to \s' = \s + a~, \quad (a= {2\pi n \over m}) ~,
\eea
where $a$ remains finite when both $n$ and $m=2N+1$ become large. However, notice that (\ref{sigma}) is not the only way to introduce the continuous parameter $\s$. In the continuum limit one may introduce a {\it local density of points} in the following way. Consider a suitable range $\dlt p$ centring around $p$ such that $\dlt p/m$ is infinitesimally small. Then the most general way of introducing the continuous parameter is
\bea
{\dlt p \over m} = \sqrt{\gamma(\s)} \dlt \s ~,
\eea
where $\dlt \s$ and $\s$ correspond to $\dlt p$ and $p$ respectively and $\sqrt{\gamma}$ is positive definite. For any other valid parametrization $\s'$ we must have,
\bea
{d\s' \over d\s} &=& \sqrt{\gamma(\s) \over \gamma'(\s')} > 0~.
\eea
This is an orientation preserving diffeomorphism of the loop, i.e. an element of $\hbox{Diff}(S^1)$. In the discussion of \S \ref{s:FNC-LM} we fixed this ambiguity by choosing $\sqrt{\gamma(\s)} = 1/2\pi$.

\subsubsection{Reparametrization isometry}
\label{ss:reparametrization}

The reparametrization invariance that arises in the continuum limit manifests itself as an isometry of the loop space. The corresponding Killing vector is given by,
\bea
\kappa^{\alpha} (\s) = \del x^{\alpha}(\s)~, \quad [\del \equiv {\del \over \del \s}]~.
\label{kappa-sigma}
\eea
In order to show that this is admitted by our large-$N$ geometry, we will prove that the following Killing equation in vielbein form \cite{tetrad} is satisfied,
\bea
\kappah^c \hat \del_c \eh^{(a)}{}_b + \hat \del_b \kappah^c \eh^{(a)}{}_c = \chi^{(a)}{}_{(d)} \eh^{(d)}{}_b ~, \quad [\hat \del_a \equiv {\del \over \del \hat z^a}] ~,
\label{killing}
\eea
where $\hat \kappa^a$ are the components of the Killing vector in (\ref{kappa-sigma}) in FNC (as constructed in \S \ref{s:FNC-LM}) and
the matrix $\chi$ satisfies,
\bea
\chi_{(ab)} + \chi_{(ba)} &=& 0~.
\label{chi}
\eea
Below we will first show that,
\bea
\kappah^{\alpha} = 0~, && \kappah^A = i \ha \yh^A~,
\label{kappah}
\eea
and then in Appendix \ref{a:killing-verification} we will verify that eq.(\ref{killing}) is indeed satisfied for $\kappah$ given above to all orders in tubular expansion.

A heuristic argument that justifies eqs.(\ref{kappah}), which was used in \cite{semi-classical}, is as follows. Recall our $T\cM$-description of loop configurations below eq.(\ref{oint-t}). According to this description and the subsequent construction of various coordinate systems,
loops can be described using the $T\cM$ coordinates as $(x^{\al}, \hat Y^{\hat \al}(\s))$. Therefore, the reparametrization Killing vector is given in this description by $\hat \kappa^{\al}(\s) = \del \hat Y^{\al}(\s)$.
Now notice that according to our construction in \S \ref{s:FNC-LM} the FNC in loop space, namely $\hat z^a = (x^{\al}, \hat y^A)$ is directly related to the above description through (\ref{yhat-Yhat}). This suggests,
\bea
\hat \kappa^{\al} &=& \oint {d\s \over 2\pi} \del \hat Y^{\al}(\s) = 0~, \quad \hat \kappa^A = \oint {d\s \over 2\pi} e^{-i\ha \s} \del \hat Y^{\hat \al}(\s) = i\ha \hat y^A~.
\label{kappa-cal}
\eea

Our approach in this paper, on the other hand, has been to understand loop space as a large-$N$ limit. Therefore a more rigorous method to obtain the above result will be to first construct a suitable vector field in the cut-off space and then take the limit. This will be discussed in Appendix \ref{a:killing-finite}.

\section{Comments and outlook}
\label{comments}

Here we make some general comments regarding the choice of the cut-off space and certain more general physical applications (besides LSQM) that the results of this paper might end up finding. 

\subsection{Choice of cut-off space}
\label{s:cut-off}

The cut-off space in eq.(\ref{cut-off}) is nothing but the total configuration space of a set of cyclically ordered 
$(2N+1)$ number of particles which may be viewed as string bits. This is similar to the usual momentum cut-off on the worldsheet where the left and right moving Fourier modes are cut-off at the value $N$. As we have seen in \S \ref{tub:LM}, these modes, along with the zero/CM mode, are related to the string bits by a discrete Fourier transformation. There is another way to see why the number of string bits is taken to be $(2N+1)$, i.e. odd. The Killing vector field corresponding to reparametrization isometry of loop space happens to vanish identically on the submanifold of vanishing loops. In \cite{kobayashi} Kobayashi proved in finite dimensional case that the space of fixed points of a continuous isometry is a submanifold which (1) has even co-dimension and (2) is totally geodesic \cite{kobayashi-nomizu}. Although Kobayashi's theorem does not strictly apply in our infinite dimensional case, but $\Dlt \hookrightarrow \cL\cM$ is a submanifold of even co-dimension in the following sense. The transverse directions are constituted by the non-zero left and right moving modes of the string and for each left moving transverse mode there exists a right moving one. Moreover the arguments of \cite{semi-classical} showed that the second fundamental form vanishes for $\Dlt \hookrightarrow \cL\cM$, making it a totally geodesic submanifold - a feature that was crucially needed to get the right form of the tachyon effective equation. At finite $N$, again Kobayashi's theorem is invalid, but this time because of a different reason: there is no continuous isometry anymore, as the entire reparametrization isometry  is replaced by the discrete isometry of cyclic permutations of the string bits. However, from the point of view of regularizing loop space quantum mechanics, it may be useful to preserve the above two features for $\Dlt \hookrightarrow \cL^{(N)}\cM$. The property of even co-dimension dictates that we choose to work with odd number of string bits and, as we have seen explicitly in \S \ref{tub:M^N},  the diagonal submanifold of $\cM^N$ is indeed a totally geodesic submanifold.

\subsection{Higher derivative corrections from finite models}
\label{s:HD}

A proper regularisation of LSQM would require a sensible finite-$N$ model to exist so that one can define cut-off versions of all possible worldsheet computations\footnote{Similar finite-$N$ string-bit models have appeared in various contexts in string theory \cite{thorn, discrete, pp-wave, ads}.}. If it is indeed possible to develop such a finite-$N$ theory, then that would be divergence-free and would describe a set of $(2N+1)$ number of interacting particles forming quantum bound states. Below we describe a line of thought with more general interests where these string inspired finite models may find applications as toy models.

Higher derivative (HD)/ $\al^{\prime}$ corrections come from the consideration of a string because it has an extended structure. This feature is independent of the nature of interactions that hold the extended body together. Therefore similar HD corrections are expected to appear in the effective theories of any other composite objects, in particular the naturally occurring ones\footnote{One may expect that this should also be true for gravitationally bound configurations, though the treatment for such classical bound configurations is expected to be very different from the quantum bound states, which is what we have in mind right now. }. Bound configurations are marked by the distinct feature that there exists an adiabatic decoupling between two different sets of degrees of freedom, namely the CM and {\it internal} degrees of freedom. These are {\it slow} and {\it fast} respectively in the Born-Oppenheimer sense. The question of interest is how to compute the HD corrections to the effective theory of the slow ones which are expected to encode the details of the {\it fast interactions}.

The approach of \cite{semi-classical} makes the above features explicit and tries to emphasise formulating a general mathematical framework (see also \cite{tubular}) of computing such corrections. One may imagine that such a framework should start from a covariant dynamical model written in positions space, i.e. the analogue of NLSM/LSQM. Then a semi-classical expansion is formulated by covariantly expanding the model around the space of all locations of the CM, which sits as a submanifold within the total configuration space. 

It is of interest to investigate whether it is indeed possible to develop such a framework for naturally occurring bound configurations. While this question may not have a straightforward answer, the aforementioned string inspired finite models may be a suitable play ground for testing/developing this mathematical framework. In this case the tubular geometry around $\Dlt \hookrightarrow {\cal L}^{(N)}\cM$, which we have computed in this work, will be of direct use.

\begin{center}
{\bf Acknowledgement}
\end{center}

I would like to thank Indranil Biswas, T. R. Ramadas and S. Ramanan for helpful discussions. I am thankful to A. P. Balachandran and Sumit R. Das for their interest and encouragement. I also thank the anonymous referees for their insightful comments and suggestions for improvements.

\appendix

\section{Tubular expansion around arbitrary submanifold}
\label{a:tubular}

Here we recall the main results of \cite{tubular}, namely the tubular expansion of vielbein around an arbitrary submanifold $\cM$ embedded in an ambient space $\cL$ of finite dimension. The Fermi normal coordinate (FNC) system is denoted by 
\bea
\hat z^a = (x^{\alpha}, \yh^A)~,  
\label{FNC}
\eea
where $x^{\alpha}$ is a general coordinate system on $\cM$. The index $A$ runs over the dimension $(\dim \cL - \dim \cM)$ of the normal space $N_x\cM$, which is taken to be arbitrary. 

We will use the following notations. Lower case symbols with a hat will be used to denote tensors of $\cL$ in FNC. Such a symbol with the argument suppressed will indicate that the tensor is being evaluated at an arbitrary point in the tubular neighborhood. For example, $\hat e^{(a)}{}_b = \hat e^{(a)}{}_b(x, \hat y)$ denote the vielbein components. An underline will be used to indicate that the quantity is being evaluated on the submanifold, e.g. $\underline{\hat e^{(a)}{}_b} = \hat e^{(a)}{}_b(x, 0)$.

The FNC expansion of vielbein is given by,
\bea
\hat e^{(a)}{}_{\beta} &=& \sum_{n \geq 0 , \{s\}_n} \lt[ \cF_{\parallel}^{(n)}(\{s\}_n) 
\underline{\hat \pi^{(a)}{}_{(b)} (\{s\}_n, \hat y) \hat e^{(b)}{}_{\beta} } 
+ \cF_{\perp}^{(n)}(\{s\}_n) \underline{\hat \pi^{(a)}{}_{(b)} (\{s\}_n, \hat y) \hat \omg_{\beta}{}^{(b)}{}_C } \hat y^C \rt] ~, \cr
\hat e^{(a)}{}_{B} &=& \sum_{n \geq 0 , \{s\}_n} \cF_{\perp}^{(n)}(\{s\}_n) \underline{\hat \pi^{(a)}{}_{(b)} (\{s\}_n, \hat y) 
\hat e^{(b)}{}_{B} } ~, 
\label{FNC-ehat}
\eea
where $\{s\}_n = \{s_1 \cdots , s_n\}$,
\bea
\cF_{\parallel}^{(n)}(\{s\}_n) &=& {C_{\parallel}^{(n)}(\{s\}) \over (s_1+ s_2+\cdots + s_n + 2n)!} ~, \cr
\cF_{\perp}^{(n)}(\{s\}_n) &=& {C_{\perp}^{(n)}(\{s\}) \over (s_1 + s_2 + \cdots + s_n + 2n +1)!}~,
\label{cF1cF3} \\
&& \cr
C_{\parallel}^{(n)}(\{s\}_n) &=& C^{s_1+s_2\cdots + s_n + 2n -2}_{s_1} C^{s_2+s_3+\cdots + s_n + 2n-4}_{s_2}  \cdots 1~, \cr
C_{\perp}^{(n)}(\{s\}_n) &=& C^{s_1+s_2\cdots + s_n + 2n -1}_{s_1} C^{s_2+s_3+\cdots + s_n + 2n-3}_{s_2}  \cdots C^{s_n+1}_{s_n} ~,
\label{C1C3}
\eea
where $C^n_r$ are binomial coefficients. 
\bea
\underline{\hat \pi (\{s\}_n, \hat y) } &=& \underline{(\hat y .\hat D^{tot})^{s_1} \hat \rho (\hat y)} 
\cdots \underline{(\hat y .\hat D^{tot} )^{s_n} \hat \rho (\hat y)}~, \cr
[\underline{(\hat y .\hat D^{tot} )^s \hat \rho (\hat y)}]^{(a)}{}_{(b)} &=& \hat y^{A^1} \cdots \hat y^{A^s} \hat y^{D} \hat y^{E} \underline{\hat D^{tot}_{A^1} \cdots \hat D^{tot}_{A^s} \hat r^{(a)}{}_{DE (b)} } ~, \cr
&=& \hat y^{A^1} \cdots \hat y^{A^s} \hat y^{D} \hat y^{E} \underline{\hat \del_{A^1} \cdots \hat \del_{A^s} \hat r^{(a)}{}_{DE (b)} }
\label{pi-hat}
\eea
where $\hat r^{(a)}{}_{bc(d)}$ is the Riemann curvature tensor, $\hat D^{tot}$ is the {\it total covariant derivative}\footnote{We follow the same definition of Riemann tensor as in \cite{tubular}. In FNC all the total covariant derivatives appearing in eqs.(\ref{pi-hat}) are same as ordinary derivatives, because all the (metric and spin) connection terms vanish. }, $\hat \del_a \equiv {\del \over \del \hat z^a}$ and 
$\hat \omega_a{}^{(b)}{}_{c} = \hat \omega_a{}^{(b)}{}_{(d)} \hat e^{(d)}{}_c$, 
$\hat \omega_a{}^{(b)}{}_{(c)}$ being the spin connection. Finally,
\bea
\underline{\hat e^{(a)}{}_{\beta}} = \dlt^a{}_{\al} \underline{\hat e^{(\alpha)}{}_{\beta}}  ~, \quad
\underline{\hat e^{(a)}{}_B} = \dlt^a_B~, 
\label{underline-e}
\eea

\section{Construction of complete coordinate transformation}
\label{a:coord-construction}

In \S \ref{s:exp-vielbein} we computed the tubular expansion of vielbein around $\Dlt \hookrightarrow \cM^N$ using the indirect method of \S \ref{s:indirect}. This method uses the general result of \cite{tubular} and the Jacobian of the coordinate transformation $\hat z \to \bar z$ evaluated at the submanifold. The latter simply follows from the general prescription of \cite{FS}. 

For a more complete understanding, here we seek to find the complete coordinate transformation $\hat z \to \bar z$ from the construction of tubular neighbourhood as described in \S \ref{tubnbh}. In Appendix \ref{a:verification} we will verify how the results of \S \ref{tub:M^N} are consistent with this construction.

The complete coordinate transformation is obtained through two steps, 
\bea
\hbox{FNC} \to \hbox{TGC} \to \hbox{DPC}~,
\label{2steps}
\eea
where the first step is to construct what we call {\it transverse general coordinates} (TGC)  which exists generically for an arbitrary tubular neighbourhood. This will be discussed in Appendix \ref{sa:arbit-TGC}. Then the second step is to perform a further coordinate transformation which is specific to our case of $\cM^N$. This will be spelled out and its correctness will be proved in Appendix \ref{sa:CT}. Some of the features of the general tubular analysis of \ref{sa:arbit-TGC} have analogues in the more familiar context of a normal neighbourhood in $\cM$. These are explained in 
Appendix \ref{sa:RNC}, which also serves the purpose of setting up notations for $\cM$-data.

\subsection{Coordinate systems in a normal neighborhood}
\label{sa:RNC}

Let $U$ be a general coordinate system in the normal neighbourhood $\cU_{\cM} \subset \cM$. The components of vielbein, metric, Christoffel symbols, covariant derivative and Riemann tensor in this system will be demote by $E^{(\al)}{}_{\beta}$, $G_{\al \beta}$, $\Gamma^{\al}{}_{\beta \gamma}$, $\nabla_{\al}$ and $R^{\al}{}_{\gamma \dlt \beta}$ respectively. As explained below eq.(\ref{spin-M}), the coordinate and non-coordinate indices will be interchanged by the vielbein. We will consider two more systems $Y'$ and $\hat Y$ to be called {\it relaxed} Riemann normal coordinate (RNC) and RNC \cite{normal} respectively where various geometric quantities will be denoted by the same symbols as above with a prime and a hat respectively.

Given a point $x \in {\cal U}_\cM$, let $\xi^{\alpha}$ be the components (in $U$-system) of an arbitrary element of 
$T_x\cM$. Then we define relaxed RNC to be,
\bea
Y'^{\alpha} &=& \xi^{\alpha}~,
\label{Y'-xi}
\eea
so that it is related to the general system in the following way,
\bea
U^{\alpha} &=& x^{\alpha} + \Exp_x^{\alpha}(Y')~,
\label{U-Y'}
\eea
where $\Exp_x: T_x\cM \to \cM$ is the exponential map in $\cM$ with origin at $x$ and it is given by,
\bea
\Exp_x^{\alpha}(\xi) &=& \xi^{\alpha} - \sum_{n\geq 0} {1\over (n+2)!} \Gamma^{\alpha}{}_{\beta^1 \cdots \beta^{n+2}} (x) \xi^{\beta^1} \cdots \xi^{\beta^{n+2}} ~,
\label{Exp}
\eea
where,
\bea
\Gamma^{\alpha}{}_{\beta^1 \cdots \beta^{n+2}} &=& \nabla_{[\beta^1} \Gamma^{\alpha}{}_{\beta^2 \cdots \beta^{n+2}]}~.
\label{multi-Gamma}
\eea
$[\cdots ]$ indicates symmetrization of indices such that,
\bea
X_{[\beta^1 \beta^2 \beta^3 \cdots]} \xi^{\beta^1} \xi^{\beta^2} \xi^{\beta^3} \cdots &=& X_{\beta^1 \beta^2 \beta^3 \cdots } \xi^{\beta^1} \xi^{\beta^2} \xi^{\beta^3} \cdots ~.
\eea
$\Gamma^{\alpha}{}_{\beta \gamma}$ and $\nabla$ are the Christoffel symbols and covariant derivative in $U$-system. Moreover, 
the covariant derivative in eq.(\ref{multi-Gamma}) acts only on the lower indices \cite{normal, gaume81}.

The expansion of vielbein in $Y'$-system is given by \cite{muller},
\bea
E'^{(\alpha)}{}_{\beta}(Y') 
&=& \sum_{n, \{s\}_n} \cF_{\perp}^{(n)}(\{s\}_n) {\Pi'_x}^{(\alpha)}{}_{(\gamma)}(\{s\}_n, Y') E'^{(\gamma)}{}_{\beta}(0) ~,
\label{E'-exp}
\eea
where $\cF_{\perp}^{(p)}(\{s\}_n)$ is given in eq.(\ref{cF1cF3}) and,
\bea
E'^{(\alpha)}{}_{\beta}(0) &=& E^{(\alpha)}{}_{\beta}(x) ~, \cr
\Pi'_x (\{s\}_n, Y') &=& (Y' . \nabla')^{s_1} \cR'_x (Y') \cdots (Y' . \nabla')^{s_p} \cR'_x (Y')~, \cr
[(Y' . \nabla')^s \cR'_x (Y')]^{(\alpha)}{}_{(\beta)} &=& Y'^{\alpha^1} \cdots Y'^{\alpha^s} Y'^{\gamma} Y'^{\dlt} \nabla'_{\alpha^1} \cdots \nabla'_{\alpha^s} R'^{(\alpha)}{}_{\gamma \dlt (\beta)}(0)~, \cr
&=& Y'^{\alpha^1} \cdots Y'^{\alpha^s} Y'^{\gamma} Y'^{\dlt} \del'_{\alpha^1} \cdots \del'_{\alpha^s} R'^{(\alpha)}{}_{\gamma \dlt (\beta)}(0)~.
\label{Pi'}
\eea
The last equality follows from the fact that all the symbols in (\ref{multi-Gamma}) vanish in $Y'$-system at $Y'=0$. The above equations can also be interpreted in general coordinates in a simple manner,
\bea
\Pi'_x (\{s\}_n, Y') &=& \Pi_x (\{s\}_n, \xi) ~,
\label{Pi'-Pi}
\eea
where the RHS is given by eqs.(\ref{Pihat}).

The reason $Y'$ is called {\it relaxed} is that the vielbein components take arbitrary values at the origin. A more standard RNC-system (as considered in \cite{muller}) $\hat Y$ is related to $Y'$ in the following way,
\bea
\Yh^{\alpha} &=& E^{(\alpha)}{}_{\beta}(x) Y'^{\beta} ~.
\label{Yh-Y'}
\eea
The expansion of the vielbein components in this system may be read directly from eq.(\ref{E'-exp}) in a straightforward manner. Each variable in that equation is replaced by the corresponding hatted one. A hatted tensor is related to the corresponding primed tensor at $U=x$ in the following way,
\bea
\hat T^{\al \beta \cdots}{}_{\gamma \dlt \cdots}(0) &=& E^{(\al)}{}_{\al'}(x) E^{(\beta)}{}_{\beta'}(x) \cdots T'{}^{\al' \beta' \cdots}{}_{\gamma' \dlt' \cdots }(0) E_{(\gamma)}{}^{\gamma'}(x) E_{(\dlt)}{}^{\dlt'}(x) \cdots ~, \cr
&=& T'{}^{(\al \beta \cdots )}{}_{(\gamma \dlt \cdots )}(0) ~.
\eea
In particular, 
\bea
\hat E^{(\alpha)}{}_{\beta} (0) &=& \dlt^{\alpha}_{\beta}~.
\label{Ehat0}
\eea
Notice that all the symbols in (\ref{multi-Gamma}) remain vanishing at the origin as expected, as they transform as tensors under 
(\ref{Yh-Y'}).

The fact that the coordinate transformation (\ref{U-Y'}) brings the vielbein components to the form given in eq.(\ref{E'-exp}) enables one to derive the following identity,
\bea
{1\over t} \del_{\xi^{\beta}} \Exp^{\gamma}_x(t\xi) E^{(\alpha)}{}_{\gamma}(x+\Exp_x(t\xi)) &=& \sum_{n, \{s\}_n} 
\cF_{\perp}^{(n)}(t, \{s\}_n) {\Pi_x}^{(\alpha)}{}_{(\gamma)}(\{s\}_n, \xi) E^{(\gamma)}{}_{\beta}(x) ~, \cr &&
\label{id-normal}
\eea
where,
\bea
\cF_{\perp}^{(n)}(t, \{s\}_n)  &=& t^{2n+s_1+\cdots +s_n} \cF_{\perp}^{(n)}(\{s\}) ~. 
\eea
The expansion of the LHS of (\ref{id-normal}) is obtained by performing ordinary Taylor expansion of $E^{(\alpha)}{}_{\beta}(U)$ around $U=x$ and using the expression (\ref{Exp}). One can check this identity order-by-order and we will use this in Appendix \ref{a:verification} to verify the analogous result for a tubular neighbourhood.

\subsection{Coordinate systems in a tubular neighborhood}
\label{sa:arbit-TGC}

Here we will perform the first step as mentioned in (\ref{2steps}). This will be done by constructing analogues of $\hat Y$, $Y'$ and $U$ systems considered in the previous sub-appendix in the context of an arbitrary tubular neighbourhood.

Given the set up of Appendix \ref{a:tubular}, the analogues of $\hat Y^{\al}$, $Y'{}^{\al}$ and $U^{\al}$ in this case are denoted as $\hat z^a=(x^{\al}, \hat y^A)$, $z'{}^a=(x^{\al}, y'{}^A)$ and $z^a=(x^{\al}, u^A)$ respectively. While the first one is FNC, the second and third may be called (by analogy with the case of normal neighbourhood) relaxed FNC and transverse general coordinates (TGC) respectively. For notations of various geometric quantities of the ambient space $\cL$ we follow similar rules as 
mentioned in the second paragraph of Appendix \ref{sa:RNC} except that here we use the corresponding lower case symbols and the covariant derivative is demoted by $\hat D_a$, $D'_a$ and $D_a$ in the above coordinate systems respectively. This is consistent with the notations already adopted in Appendix \ref{a:tubular}. We will also continue to use the rule for describing the argument of geometric quantities as mentioned in the second paragraph of Appendix \ref{a:tubular}. 

The coordinate transformations,
\bea
\zh^a = (x^{\alpha}, \yh^A) \to z'^a = (x^{\alpha}, y'^A) \to z^a=(x^{\alpha}, u^A),
\label{CT-step1}
\eea
are given by,
\bea
\yh^A &=& \underline{e'^{(A)}{}_B} y'^B~,
\label{zh-z'}
\eea
and
\bea
u^A &=& \exp^A(x,y')  ~, \cr
&:=& y'^A - \sum_{n\geq 0} {1\over (n+2)!} \underline{\gamma^A{}_{B^1 \cdots B^{n+2}}} y'^{B^1} \cdots y'^{B^{n+2}} ~, 
\label{u-A} \\
\gamma}^A{}_{B^1 \cdots B^{n+2} &=& D_{[B^1} {\gamma}^A{}_{B^2 \cdots B^{n+2}]} ~.
\label{multi-gamma}
\eea
We now explain the above equations. In FNC, the transverse vielbein at the submanifold is given by identity matrix (see the last equation in (\ref{underline-e})). In relaxed FNC, it is given by $\underline{e'{}^{(A)}{}_B}$, which is arbitrary. TGC is related to this one through the exponential map of the ambient space along the directions transverse to the submanifold. Just like in (\ref{multi-Gamma}), the covariant derivative in eq.(\ref{multi-gamma}) acts only on the lower indices. 

Given that TGC and relaxed FNC are same up to linear order from the submanifold, components of any tensor are identical in these two systems at the submanifold. In particular,
\bea
\underline{e^{(A)}{}_B} &=& \underline{e'{}^{(A)}{}_B} ~.
\label{e-e'}
\eea

We now derive the analogue of eq.(\ref{id-normal}). To this end let us first consider an element of the normal space $N_x\cM$. We denote its components in $\hat z$, $z'$ and $z$-systems by $\hat \xi^A$, $\xi'{}^A$ and $\xi^A$ respectively. The above coordinate transformations imply,
\bea
\hat \xi^A &=& \underline{e^{(A)}{}_B} \xi'{}^B ~, \quad \xi'{}^A = \xi^A~. 
\label{xihat-xiprime-xi}
\eea
Let us now consider relating the vielbein components in TGC and FNC using the Jacobaian matrix of the above coordinate transformations. Since the vielbein components in FNC can be expanded in terms of 
$\hat y^A = \hat \xi^A$ following the results of Appendix \ref{a:tubular}, it should be possible to express the same in TGC in terms of the tubular expansion coefficients and the Jacobian matrix which involves both tangential and transverse derivatives of the exponential maps in $\cL$. The precise forms of these relations are as follows,
\bea
&& {1\over t} \del_{\xi^B} \exp^{C}(x, t\xi) e^{(a)}{}_{C}(x, \exp(x, t\xi)) = \sum_{n, \{s\}_n} \cF_{\perp}^{(n)}(t, \{s\}_n) \underline{\pi^{(a)}{}_{C}(\{s\}_n; x, \xi) e^{(C)}{}_{B} }~, \cr &&
\label{id-tubular-prp} \\
&& e^{(a)}{}_{\beta}(x, \exp(x,t\xi)) + \del_{x^\beta} \exp^{C}(x, t\xi) e^{(a)}{}_{C}(x, \exp(x,t\xi)) \cr
&=& \sum_{n \geq 0 ,\{s\}_n} \lt[ \cF^{(n)}_{\parallel}(t, \{s\}_n) \underline{\pi^{(a)}{}_{(b)} (\{s\}_n, \xi) 
e^{(b)}{}_{\beta} }  \rt. \cr
&& \lt. 
+  t \cF_{\perp}^{(n)}(t, \{s\}_n) \lt( \underline{\pi^{(a)}{}_{(b)} (\{s\}_n, \xi) 
\omg_{\beta}{}^{(b)}{}_C }
+ \underline{\pi^{(a)}{}_{(B)}  (\{s\}_n, \xi) \del_{x^\beta} 
e^{(B)}{}_{C} } \rt)  \xi^{C} \rt] ~, \cr &&
\label{id-tubular-prl}
\eea
where,
\bea
\cF^{(n)}_{\parallel}(t, \{s\}_n) &=& t^{2n+s_1+\cdots +s_n} \cF^{(n)}_{\parallel}(\{s\}_n) ~. 
\eea
The matrix $\pilb (\{s\}_n, \xi)$ has been defined in eqs.(\ref{pi-hat}) in terms of quantities in FNC. But for our purpose here it will be useful to interpret it in TGC,
\bea
\underline{\pi (\{s\}_n, \xi) } &=& \underline{(\xi . D^{tot} )^{s_1} \rho (\xi)} \cdots \underline{(\xi . D^{tot} )^{s_n} \rho (\xi)}~, \cr
[\underline{(\xi . D^{tot} )^s \rho (\xi)}]^{(a)}{}_{(b)} &=& \xi^{A^1} \cdots \xi^{A^s} \xi^{B} \xi^{C} 
\underline{D^{tot}_{A^1} \cdots D^{tot}_{A^s} r^{(a)}{}_{BC (b)}} ~.
\eea
This way we write both the equations (\ref{id-tubular-prp}, \ref{id-tubular-prl}) entirely in terms of TGC. While eq.(\ref{id-tubular-prp}) has the same content as the one in (\ref{id-normal}), eq.(\ref{id-tubular-prl}) arises due to the fact that now we are dealing with a higher dimensional submanifold rather than a point. Notice the appearance of the derivative of exponential map with respect to base point on the LHS. 
This is responsible for giving rise to curvature terms with the right coefficients on the RHS. This fact is very crucial in the verification of our results in Appendix \ref{a:verification}. Below we explicitly verify eq.(\ref{id-tubular-prl}) up to second order in $t$. 

To this end we first compute ordinary Taylor expansion in general coordinates. The LHS of (\ref{id-tubular-prl}) upto quadratic order is found to be,
\bea
\hbox{LHS}^{(2)} &=& \underline{e^{(a)}{}_{\beta} } + t \xi^C \underline{e^{(a)}{}_{\beta,C}} 
+ {t^2 \over 2} \xi^{C_1} \xi^{C_2} (\underline{e^{(a)}{}_{\beta,C_1 C_2}}  - \underline{\gamma^C_{C_1 C_2} e^{(a)}{}_{\beta,C} } \cr
&& - \del_{\beta} (\underline{\gamma^C{}_{C_1 C_2} } ) \underline{e^{(a)}{}_C} ) ~,
\label{LHS2}
\eea
where we have used the following notations $\underline{f_{,A} } \equiv \lim_{u\to 0} \del_{u^A} f$ and $\del_{\alpha} f \equiv \del_{x^{\alpha}} f $. By manipulating the RHS, one finds up to second order,
\bea
\hbox{RHS}^{(2)} &=& \underline{e^{(a)}{}_{\beta} } + t \xi^C (\underline{\omg_{\beta}{}^{(a)}{}_C } 
+ \dlt^{a}{}_{B} \del_{\beta} \underline{e^{(B)}{}_C} ) + {t^2 \over 2} {\rho}^{(a)}{}_{(b)}(x,\xi) 
\underline{e^{(b)}{}_{\beta} } ~.
\label{RHS2}
\eea
Comparing (\ref{LHS2}) and (\ref{RHS2}) we get,
\bea
\xi^C \underline{e^{(a)}{}_{\beta,C} } &=& \xi^C (\dlt^{(a)}{}_{(B)} \del_{\beta} \underline{e^{(B)}{}_C } 
+ \underline{\omg_{\beta}{}^{(a)}{}_C } ) ~, 
\label{verify1} \\
 {\rho}^{(a)}{}_{(b)}(x,\xi) \underline{e^{(b)}{}_{\beta} } &=&
\xi^{C_1} \xi^{C_2} (\underline{e^{(a)}{}_{\beta,C_1 C_2} } - \underline{\gamma^C_{C_1 C_2} e^{(a)}{}_{\beta,C} } - \underline{ \del_{\beta} ( \gamma^C{}_{C_1 C_2} ) e^{(a)}{}_C} ) ~.
\label{verify2}
\eea
These equations are verified below.

\subsubsection{Proof of equations (\ref{verify1}, \ref{verify2})}
\label{ssa:arbit-proof}

We first prove eq.(\ref{verify1}). Using the vanishing of {\it total covariant derivative} (along transverse to the submanifold) we write,
\bea
\xi^C \underline{e^{(a)}{}_{\beta, C} } &=& \xi^C \underline{\gamma^d_{\beta C} e^{(a)}{}_d} 
- \xi^C \underline{\omg_C{}^{(a)}{}_{(b)} e^{(b)}{}_{\beta} } ~.
\eea
Due to the special property of FNC, $\hat y^C \hat \omg_C{}^{(a)}{}_{(b)}$ vanishes everywhere \cite{tubular}. This implies,
\bea
0=\hat \xi^C \underline{\hat \omg_C{}^{(a)}{}_{(b)}} = \xi^C \underline{\omg_C{}^{(a)}{}_{(b)} }
\label{xi-omega}
\eea
Therefore,
\bea
\hbox{LHS of (\ref{verify1})} = \xi^C \underline{e^{(a)}{}_{\beta, C} } &=& 
\xi^C \underline{\gamma^d_{\beta C} e^{(a)}{}_d } ~.
\label{LHS-verify1}
\eea
Actually the above identity holds true with $\beta$ replaced by $b$. 
\bea
\xi'^C \underline{e^{(a)}{}_{b, C} } &=& \xi'^C \underline{\gamma'{}^d_{b C} e'{}^{(a)}{}_d}  ~.
\label{elb-der}
\eea

To compute the RHS of (\ref{verify1}) we use vanishing of total covariant derivative along the submanifold,
\bea
\del_{\beta} e^{(a)}{}_C - \gamma^d_{\beta C} e^{(a)}{}_d + \omg_{\beta}{}^{(a)}{}_{(b)} e^{(b)}{}_C = 0~.
\eea
Using the fact that $e^{(a)}{}_b$ is block diagonal everywhere on the submanifold, one can re-write the above equation as,
\bea
\xi^C(\dlt^a{}_B \del_{\beta} \underline{e^{(B)}{}_C } + \underline{\omg_{\beta}{}^{(a)}{}_C} ) &=& \xi^C \underline{\gamma^d_{\beta C} e^{(a)}{}_d } ~,
\eea
which is the RHS of (\ref{verify1}) and is same as the result in (\ref{LHS-verify1}).

To prove eq.(\ref{verify2}) we proceed as follows. We compute total covariant derivative twice along transverse to the submanifold. The result is,
\bea
D^{tot}_{C_1} D^{tot}_{C_2} e^{(a)}{}_{\beta}
&=& e^{(a)}{}_{\beta, C_1 C_2} - \del_{C_1} \gamma^b_{C_2 \beta} e^{(a)}{}_b - \gamma^b_{C_2 \beta} e^{(a)}{}_{b, C_1} + D_{C_1} \omg_{C_2}{}^{(a)}{}_{(d)} e^{(d)}{}_{\beta} \cr
&& - \gamma^b_{C_1 C_2} e^{(a)}{}_{\beta, b} + \gamma^b_{C_1 C_2} \gamma^d_{b \beta} e^{(a)}{}_d - \gamma^b_{C_1 \beta} e^{(a)}{}_{b,C_2} + \gamma^b_{C_1 \beta} \gamma^d_{C_2 b} e^{(a)}{}_d \cr
&& - \gamma^b_{C_1 \beta} \omg_{C_2}{}^{(a)}{}_{(d)} e^{(d)}{}_b + \omg_{C_1}{}^{(a)}{}_{(b)} e^{(b)}{}_{\beta, C_2} - \gamma^d_{C_2 \beta} \omg_{C_1}{}^{(a)}{}_{(b)} e^{(b)}{}_d \cr
&& + \omg_{C_1}{}^{(a)}{}_{(b)} \omg_{C_2}{}^{(b)}{}_{(d)} e^{(d)}{}_{\beta} ~, \cr
&=& 0~.
\label{double-cov-der}
\eea
Due to (\ref{xi-omega}), the last four terms do not contribute in $\xi^{C_1} \xi^{C_2} \underline{D^{tot}_{C_1} D^{tot}_{C_2} e^{(a)}{}_{\beta}}$. Below we compute the contribution of the fourth term.
\bea
\xi^{C_1} \xi^{C_2} \underline{D_{C_1} \omg_{C_2}{}^{(a)}{}_{(b)}} &=& \hat \xi^{C_1} \hat \xi^{C_2} \underline{\hat D_{C_1} \hat \omg_{C_2}{}^{(a)}{}_{(b)}} 
= \hat \xi^{C_1} \hat \xi^{C_2} \underline{\hat \del_{C_1} \hat \omg_{C_2}{}^{(a)}{}_{(b)}} = 0~.
\label{xi2-D-omega}
\eea
The last equality can be shown as follows,
\bea
0 &=& \hat y^{C_1} \hat \del_{C_1} \lt[ \hat y^{C_2} \hat \omg_{C_2}{}^{(a)}{}_{(b)} \rt] =
\hat y^A \hat \omg_A{}^{(a)}{}_{(b)} + \hat y^{C_1} \hat y^{C_2} \hat \del_{C_1} \hat \omg_{C_2}{}^{(a)}{}_{(b)} = \hat y^{C_1} \hat y^{C_2} \hat \del_{C_1} \hat \omg_{C_2}{}^{(a)}{}_{(b)}~. \cr &&
\eea
Therefore,
\bea
\hbox{RHS of (\ref{verify2})} &=& \xi^{C_1} \xi^{C_2} \elb^{(a)}{}_b \lt[ \underline{ \del_{C_1} \gamma^b_{C_2 \beta} }
- \underline{ \del_{\beta} \gamma^b_{C_1 C_2} } + \underline{ \gamma^d_{C_2 \beta} \gamma^b_{C_1 d} }
- \underline{ \gamma^d_{C_1 C_2} \gamma^b_{d \beta} } \rt]~, 
\eea
which is precisely the LHS of (\ref{verify2}). To get to the first line we have used (\ref{elb-der}) and the fact that, 
\bea
\underline{\gamma^{\alpha}_{BC} } &=& 0 ~, 
\eea 
which can in tern be justified as follows,
\bea
\underline{\gamma^{\alpha}_{BC} } &=& \underline{\gamma'^{\alpha}_{BC} } ~, \quad \lt[\hbox{because, } \underline{\lt(\del x^{\alpha} \over \del z'^d\rt) \lt(\del^2 z'^d \over \del z^B \del z^C \rt)} =  0 \rt] ~, \cr
&=& \underline{e^{(D)}{}_B e^{(E)}{}_C \hat \gamma^{\alpha}_{DE} }
= 0~.
\eea

\subsection{Complete coordinate transformation}
\label{sa:CT}

Here we perform the second step in (\ref{2steps}). We begin by recalling the first equation in (\ref{xp-xip}) which is given in terms of exponential map and tangent vector in $\cM$. Therefore the complete coordinate transformation $\hat z \to z \to \bar z$, when expressed entirely in terms of $\cM$-data, should reduce to this equation. Below we will first construct $z \to \bar z$ and then show that this is indeed the case.

The transformation from TGC to DPC is given by introducing a new coordinate system $\tilde z^a$ as an intermediate step,
\bea
z^a=(x^{\al}, u^A) \to \tilde z^a =(\tilde x^{\al},  \tilde u^A) \to \bar z^{\bar a} = (\{x_p^{\al_p}\})  ~,
\label{CT-step2}
\eea
where,
\bea
x_p^{\alpha_p} &=& \sqrt{N} (R^T)^{\alpha_p}{}_b \tilde z^b~, 
\label{zb-ztilde} \\
\tilde x^{\alpha} &=& \tilde{\exp}_{\parallel}^{\alpha}(x, \tilde{\log}_{\perp} (x, u)) ~, \quad 
\tilde u^A = u^A~.
\label{ztilde-z}
\eea
The functions $\tilde{\exp}_{\parallel}$ and $\tilde{\log}_{\perp}$ are defined as follows. Let us consider the transverse vector 
$(t_1, t_2, \cdots , t_N) \in T_{(x, x, \cdots ,x)}\cM^N$ satisfying eq.(\ref{sum-tp}). Its components are given by $\bar \xi^{\bar a}$ in DPC (see eq.(\ref{xi-transverse})) and by $(\hat \xi^{\al}=0~, \hat \xi^A)$, $(\xi'{}^{\al}=0~, \xi'{}^A)$ and $(\xi^{\al}=0~, \xi^A)$ in FNC, relaxed FNC and TGC respectively (see near eqs.(\ref{xihat-xiprime-xi})). Furthermore, we denote the components of the same vector in $\tilde z$-system by $(\tilde \xi^{\al}=0~, \tilde \xi^A)$. Their inter-relations are given by eqs.(\ref{xihat-xiprime-xi}) and (following eqs.(\ref{zb-ztilde}, \ref{ztilde-z})),
\bea
\tilde \xi^A = \xi^A = {1\over \sqrt{N}} R^A{}_{\bar b} \bar \xi^{\bar b}~.
\label{xitilde-xibar}
\eea
Given this, we then define the ({\it transverse}) exponential map in $\tilde z$-system as,
\bea
\tilde u^A = \tilde{\exp}_{\perp}^A (x, \tilde \xi) &=& \tilde \xi^A - \sum_{n\geq 0} {1\over (n+2)!} \underline{\tilde \gamma^A{}_{B^1 \cdots B^{n+2}}} \tilde \xi^{B^1} \cdots \tilde \xi^{B^{n+2}} ~,
\label{exp-tilde-A}
\eea
where the coefficients are given by the same expression as in (\ref{multi-gamma}), but now with reference to $\tilde z$-system. The above map can be inverted order-by-order within the tubular neighbourhood. We denote this inverted map as,
\bea
\tilde \xi^A &=& \tilde{\log}_{\perp}^A (x, \tilde u)~.
\label{xitilde-utilde}
\eea 
Finally, the exponential map with a longitudinal index as appearing in the first equation in (\ref{ztilde-z}) is defined by,
\bea
\tilde{\exp}_{\parallel}^{\alpha}(x, \tilde \xi) &=& x^{\alpha} - \sum_{n\geq 0} {1\over (n+2)!} \underline{\tilde \gamma^{\alpha}{}_{B^1 \cdots B^{n+2}}} \tilde \xi^{B^1} \cdots \tilde \xi^{B^{n+2}} ~,
\label{exp-tilde-alpha}
\eea
where the coefficients are given by the same as appearing in (\ref{exp-tilde-A}) with the transverse index $A$ replaced by the longitudinal one $\alpha$. By this we finish specifying the complete coordinate transformation.

We now prove that the coordinate transformations in (\ref{zb-ztilde}, \ref{ztilde-z}) are indeed the right ones.
To this end we notice that because of the second equation in (\ref{ztilde-z}), we must have,
\bea
\underline{\tilde \gamma^A{}_{BC \cdots}} &=& \underline{\gamma^A{}_{BC \cdots}} ~.
\eea
Furthermore, because of the coordinate transformation in (\ref{zb-ztilde}), we must also have,
\bea
\underline{\tilde \gamma^a{}_{bc \cdots }} &=& [{1\over \sqrt{N}} R^a{}_{\bar a} ] 
[\sqrt{N} (R^{-1})^{\bar b}{}_b] [\sqrt{N} (R^{-1})^{\bar c}{}_c ] \cdots \underline{\bar \gamma^{\bar a}{}_{\bar b \bar c \cdots } }~.
\eea
Since the symbols $\bar \gamma^{\bar a}{}_{\bar b \bar c \cdots}$ (see eq.(\ref{multi-gamma})) are constructed out of Christoffel symbols and their derivatives, the Weyl weight $w$ (as defined in eq.(\ref{weyl-weight})), for such quantites is zero. Therefore we must have, 
\bea
\underline{\bar \gamma_p^{\alpha_p}{}_{\beta_q \xi_r \cdots } } = \Gamma^{\alpha_p}{}_{\beta_q \xi_r \cdots}(x) \dlt_{p,q} \dlt_{q, r} \cdots ~. 
\eea

This enables us to write the coordinate transformation $z \to \tilde z \to \zb$ entirely in terms of $\cM$-data,
\bea
x_p^{\alpha_p} &=& \sqrt{N} (R^T)^{\alpha_p}{}_{\beta} x^{\beta} + \sqrt{N} (R^T)^{\alpha_p}{}_B \tilde \xi^B
- \sqrt{N} (R^T)^{\alpha_p}{}_b \sum_{n \geq 0} {1\over (n+2)!} \underline{\tilde \gamma^b{}_{C^1 \cdots C^{n+2}}} \tilde \xi^{C^1} \cdots \tilde \xi^{C^{n+2}}~,\cr
&=& x^{\alpha_p} + \xi_p^{\alpha_p} - \sum_{n \geq 0} {1\over (n+2)!}
\Gamma^{\alpha_p}{}_{\beta_p^1 \cdots \beta_p^{n+2}}(x) \xi_p^{\beta_p^1} \cdots \xi_p^{\beta_p^{n+2}}~, \cr
&=& x^{\alpha_p} + \Exp_x^{\al_p}(\xi_p) ~,
\eea
which is precisely the first equation in (\ref{xp-xip}).

\section{Verification of results}
\label{a:verification}

Equations (\ref{ehat0-alpha}, \ref{ehat0-A}) and (\ref{pihat-lowerbar}), along with the results summarized in Appendix \ref{a:tubular}, give the tubular expansion of vielbein components in FNC around $\Dlt \hookrightarrow \cM^N$ written entirely in terms of $\cM$-data. Note that this expansion is given in terms of $\xi_p$, as defined below eq.(\ref{t-vector}). Also recall that it was obtained by using an indirect method where only a limited information of the Jacobian matrix of the relevant coordinate transformation was used - namely its value restricted to the submanifold. 

On the other hand, in the previous appendix we constructed the complete coordinate transformation as given by eqs.(\ref{CT-step1}) and (\ref{CT-step2}). This gives the Jacobian matrix everywhere and therefore can in principle be used to compute the aforementioned tubular expansion using direct method. The goal of this Appendix is to show, up to quadratic order, that this method indeed gives the same result as obtained by using the indirect method. This gives evidence for overall consistency of all our results. 

Our analysis below is divided into two parts. In \S \ref{sa:Jacobian} we show that the Jacobian matrix $\underline{J}$ constructed and used in \S \ref{s:indirect} is compatible with the coordinate transformations (\ref{CT-step1}) and (\ref{CT-step2}). Then in \S \ref{sa:direct} we formulate and then verify the consistency equations.

\subsection{Jacobian matrix at submanifold}
\label{sa:Jacobian}

To show that the Jacobian matrix computed using eqs.(\ref{CT-step1}) and (\ref{CT-step2}), restricted to the submanifold, indeed gives the results in (\ref{underline-J}-\ref{calE}), we need to compute,
\bea
\underline{J^{\al}{}_{\bar b}} &=& \underline{ \lt({\del \hat z^{\al} \over \del {z'}^{c'}}\rt) \lt({\del {z'}^{c'} \over \del z^c }\rt) \lt({\del z^c \over \del {\tilde z}^{\tilde c}}\rt) \lt({\del {\tilde z}^{\tilde c} \over \del {\bar z}^{\bar b} }\rt) } ~,
\label{verify-underline-J-al} \\
\underline{J^A{}_{\bar b}} &=& \underline{ \lt({\del \hat z^A \over \del {z'}^{c'}}\rt) \lt({\del {z'}^{c'} \over \del z^c }\rt) \lt({\del z^c \over \del {\tilde z}^{\tilde c}}\rt) \lt({\del {\tilde z}^{\tilde c} \over \del {\bar z}^{\bar b} }\rt) } ~.
\label{verify-underline-J-A}
\eea

We first verify eq.(\ref{verify-underline-J-al}). It is obvious from eq.(\ref{CT-step1}) that\footnote{We have in mind the following notation for example: $z^c = (z^{\xi}, z^C)$, which associates $c$ with $\xi$.} $\underline{ \lt({\del \hat z^{\al} \over \del {z'}^{c'}}\rt) } = ( \dlt^{\al}{}_{\xi'}, 0)$ and similarly for the second factor. It turns out that for the third factor also one has: $\underline{\lt(\del z^{\xi} \over \del {\tilde z}^{\tilde c} \rt) } = (\dlt^{\xi}{}_{\tilde \xi}, 0)$. This is shown by using eqs.(\ref{exp-tilde-alpha}, \ref{xitilde-utilde}) and (\ref{exp-tilde-A}) in eq.(\ref{ztilde-z}) to argue that: $x^{\xi} = \dlt^{\xi}{}_{\tilde \xi} \tilde x^{\tilde \xi} + O(u^2)$. Finally it remains to use eq.(\ref{zb-ztilde}) to show that the correct result is reproduced. 

We now consider the second equation (\ref{verify-underline-J-A}). Again using eqs.(\ref{CT-step1}) and 
(\ref{e-e'}) it is clear that the first two factors give: $\underline{\lt( {\del \hat z^A \over \del {z'}^{c'}}\rt) \lt({\del {z'}^{c'} \over \del z^c } \rt)} = (0, \underline{e^{(A)}{}_C})$. Then using argument similar to above for (\ref{CT-step2}) one shows that,
\bea
\underline{J^A{}_{\beta_p}} &=& {1\over \sqrt{N}} \underline{e^{(A)}{}_B} R^B{}_{\beta_p}~.  
\eea
Therefore, according to the second equation in (\ref{underline-J}), we must have as matrices,
\bea
\underline{e} &=& {\cal E}(x)~,
\label{e-calE}
\eea
where ${\cal E}(x)$ is as given in eqs.(\ref{calE-def}, \ref{calE}). To show this we proceed as follows. According to the first equation in (\ref{underline-J}) (which has already been argued in the previous paragraph) and the above equation, we must have
$\underline{K^{\al_p}{}_B} = \sqrt{N} (R^T)^{\al_p}{}_C \underline{e_{(B)}{}^C}$.
Now noting the relation,
$\underline{\hat e^{(A)}{}_b} = R^A{}_{\bar a} \underline{\bar e^{(\bar a)}{}_{\bar b} K^{\bar b}{}_b }$
which is obtained by combining the second equations in (\ref{ebar-prl-perp}) and (\ref{ehat-ebar}), and using eqs.(\ref{underline-e}) and (\ref{ebar}), one gets: $\sum_p R^A{}_{\al_p} E^{(\al_p)}{}_{\beta_p}(x) (R^T)^{\beta_p}{}_C \underline{e_{(B)}{}^C} = \dlt^A{}_B$. This equation, given the value of $R$ as given in (\ref{R-def}), implies (\ref{e-calE}). This establishes the fact that the matrix $\underline{J}$ constructed in \S \ref{s:indirect} is compatible with the coordinate transformation considered in (\ref{CT-step1}, \ref{CT-step2}).

\subsection{Verification using direct method}
\label{sa:direct}

Here we first formulate the actual equations implied by consistency with direct method and then verify those equations up to second order. All quantities to be evaluated in the neighbourhood are to be viewed as expansions in powers of $\xi_p$.

The equation that relates vielbein components in FNC and DPC is given by,
\bea
\hat e^{(a)}{}_b &=& R^a{}_{\bar a} \bar e^{(\bar a)}{}_{\bar b} K^{\bar b}{}_b ~.
\label{ehat-ebar}
\eea
We first compute the LHS of (\ref{ehat-ebar}) up to second order in $\xi_p$. Using eqs.(\ref{ehat0-alpha}, \ref{ehat0-A}) and (\ref{pihat-lowerbar}) along with the results summarized in Appendix \ref{a:tubular}, one first calculates up to second order, 
\bea
\underline{\hat \pi ^{(\al)}{}_{(c)}} (\{0\}_1, \hat y) \hat e_0^{(c)}{}_{\beta} &=& 
{1\over N} R^{(\al)}{}_{\gamma \dlt \beta}(x) \sum_p \xi_p^{\gamma} \xi_p^{\dlt} + \cdots ~, \cr
\underline{\hat \pi ^{(A)}{}_{(c)}} (\{0\}_1, \hat y) \hat e_0^{(c)}{}_{\beta} &=& {1\over \sqrt{N} } R^{(\hat \al)}{}_{\gamma \dlt \beta}(x) \sum_p O_{\ha p} \xi_p^{\gamma} \xi_p^{\dlt} + \cdots ~, \cr
\underline{\hat \pi ^{(\al)}{}_{(c)}} (\{0\}_1, \hat y) \hat e_0^{(c)}{}_B 
&=& {1\over \sqrt{N} } R^{(\al)}{}_{\gamma \dlt (\hat \beta) }(x) \sum_p O_{\hb p} \xi_p^{\gamma} \xi_p^{\dlt} + \cdots ~,  \\ && \cr
\underline{\hat \pi ^{(A)}{}_{(c)}} (\{0\}_1, \hat y) \hat e_0^{(c)}{}_B &=& 
R^{(\hat \al)}{}_{\gamma \dlt \hat \beta}(x) \sum_p O_{\ha p} O_{\hb p} \xi_p^{\gamma} \xi_p^{\dlt} 
+ \cdots ~.
\eea
Using this expansions one gets the following final results for the LHS of (\ref{ehat-ebar}),
\bea
\hat e^{(\al)}{}_{\beta} &=& E^{(\al)}{}_{\beta}(x) + \displaystyle{ {1\over 2 N} R^{(\al)}{}_{\gamma \dlt \beta}(x) \sum_p \xi_p^{\gamma} \xi_p^{\dlt}} + O(\xi^3) ~, \cr 
\hat e^{(\al)}{}_B &=& {1\over 6 \sqrt{N} } R^{(\al)}{}_{\gamma \dlt (\hat \beta) }(x) \sum_p O_{\hb p} \xi_p^{\gamma} \xi_p^{\dlt} + O(\xi^3)~, \cr
\hat e^{(A)}{}_{\beta} &=& \displaystyle{{1\over \sqrt{N}} \Omega_{\beta}{}^{(\hat \alpha)}{}_{\gamma}(x) \sum_p O_{\ha p} \xi^{\gamma}_p }  + \displaystyle{ {1\over 2 \sqrt{N} } R^{(\hat \al)}{}_{\gamma \dlt \beta}(x) \sum_p O_{\ha p} \xi_p^{\gamma} \xi_p^{\dlt}}   + O(\xi^3) ~, \cr
\hat e^{(A)}{}_B &=& \dlt^A{}_B + {1\over 6} R^{(\hat \al)}{}_{\gamma \dlt \hat \beta}(x) \sum_p O_{\ha p} O_{\hb p} \xi_p^{\gamma} \xi_p^{\dlt} + O(\xi^3) ~.
\label{verify-LHS}
\eea

We will now proceed to compute the RHS of (\ref{ehat-ebar}). To find it as an expansion in $\xi_p$, we first write it as an expansion in $\tilde \xi$ and then use the relation (\ref{xitilde-xibar}). To this end we write the RHS of (\ref{ehat-ebar}) in $\tilde z$-frame, 
\bea
\cQ^{(a)}{}_b &:=& R^a{}_{\bar a} \bar e^{(\bar a)}{}_{\bar b} K^{\bar b}{}_b
= \tilde e^{(a)}{}_d (\tilde{\exp}_{\parallel}(x, \tilde \xi), \tilde{\exp}_{\perp}(x, \tilde \xi)) \lt(\del \tilde z^d \over \del \hat z^b \rt) ~, 
\label{calQ-def}
\eea
where the second factor should be understood as,
\bea
\lt(\del \tilde z^{\dlt} \over \del \hat z^{\beta} \rt) &=& \del_{\beta} \tilde{\exp}^{\dlt}_{\parallel}(x, \tilde \xi (\hat y)) =  \del_{\beta} \tilde{\exp}^{\dlt}_{\parallel}(x, \tilde \xi) + (\del_{\beta} \underline{e_{(E)}{}^D}) \underline{e^{(E)}{}_F} \tilde \xi^F \tilde{\del}_D \tilde{\exp}^{\dlt}_{\parallel}(x, \tilde \xi) ~, \cr
\lt(\del \tilde z^{\dlt} \over \del \hat z^B \rt) &=& \hat \del_B \tilde{\exp}^{\dlt}_{\parallel}(x, \tilde \xi (\hat y)) =  \underline{e_{(B)}{}^D} \tilde{\del}_D \tilde{\exp}_{\parallel}^{\dlt} (x, \tilde \xi) ~, \cr
\lt(\del \tilde z^D \over \del \hat z^{\beta} \rt) &=& \del_{\beta} \tilde{\exp}^D_{\perp}(x, \tilde \xi (\hat y))
=  \del_{\beta} \tilde{\exp}^D_{\perp}(x, \tilde \xi) + (\del_{\beta} \underline{e_{(E)}{}^D}) \underline{e^{(E)}{}_F} \tilde \xi^F \tilde{\del}_D \tilde{\exp}^D_{\perp}(x, \tilde \xi) ~,  \cr
\lt(\del \tilde z^D \over \del \hat z^B \rt) &=& \hat \del_B \tilde{\exp}^D_{\perp}(x, \tilde \xi (\hat y)) =  \underline{e_{(B)}{}^D} \tilde{\del}_D \tilde{\exp}_{\perp}^D (x, \tilde \xi) ~,
\eea
where we have used the following notations: $\del_{\beta} \equiv {\del \over \del x^{\beta}}, \tilde{\del}_B \equiv {\del \over \del \tilde \xi^B},  \hat{\del}_B \equiv {\del \over \del \hat y^B} $ and,
\bea
\tilde \xi^A(\hat y) &=& \underline{e_{(B)}{}^A} \hat y^B~, 
\eea
which is obtained by using eqs.(\ref{xitilde-xibar}, \ref{yh-xibar}). In order to obtain ${\cal Q}$ as a power series in $\tilde \xi$, one performs ordinary Taylor expansion of all the relevant quantities around $\tilde \xi=0$. Using such Taylor expansions we will show in \S \ref{ssa:verify-proof} that,
\bea
{\cal Q}^{(a)}{}_{\beta} &=& \underline{\tilde e^{(a)}{}_{\beta}} + \underline{\tilde \omg_{\beta}{}^{(a)}{}_C} \tilde{\xi}^C 
+ {1\over 2} \underline{\tilde r^{(a)}{}_{C^1 C^2 \beta} } \tilde{\xi}^{C^1} \tilde{\xi}^{C^2} + O(\tilde \xi^3)  ~, 
\label{calQ-beta} \\
{\cal Q}^{(a)}{}_B &=& \underline{e_{(B)}{}^D \tilde e^{(a)}{}_D} 
+ {1\over 6} \underline{e_{(B)}{}^D \tilde r^{(a)}{}_{C^1 C^2 D} } \tilde{\xi}^{C^1} \tilde{\xi}^{C^2} + (\tilde \xi^3) ~.
\label{calQ-B}
\eea
We now express the RHS of the above equations back in terms of DPC,
\bea
{\cal Q}^{(a)}{}_{\beta} &=& \sqrt{N} R^a{}_{\bar a} (\underline{\bar e^{(\bar a)}{}_{\bar b}} + \underline{\bar \omg_{\bar b}{}^{(\bar a)}{}_{\bar c} } \bar \xi^{\bar c} + {1\over 2} \underline{\bar r^{(\bar a)}{}_{\bar c^1 \bar c^2 \bar b} } \bar \xi^{\bar c^1} \bar \xi^{\bar c^2} + O(\bar \xi^3)  )
(R^T)^{\bar b}{}_{\beta} ~, \cr
{\cal Q}^{(a)}{}_B &=& \underline{e_{(B)}{}^D} \lt\{ \sqrt{N} R^a{}_{\bar a} (\underline{\bar e^{(\bar a)}{}_{\bar b}} + {1\over 6} \underline{\bar r^{(\bar a)}{}_{\bar c^1 \bar c^2 \bar b } } \bar \xi^{\bar c^1} \bar \xi^{\bar c^2}  + O(\bar \xi^3) ) (R^T)^{\bar b}{}_D \rt\}~.
\eea
These expressions can in turn be evaluated in terms of $\cM$-data to get the same results as in (\ref{verify-LHS}).

\subsubsection{Proof of equations (\ref{calQ-beta}, \ref{calQ-B})}
\label{ssa:verify-proof}

Results for the ordinary Taylor expansions up to quadratic order relevant to the computation of (\ref{calQ-def}) are given by,
\bea
&& \del_{\beta} \tilde{\exp}_{\parallel}^{\dlt} (x, \tilde \xi) \tilde{e}^{(a)}{}_{\dlt} (\tilde{ \exp}_{\parallel}(x, \tilde \xi), \tilde{\exp}_{\perp}(x, \tilde \xi)) ~, \cr
&=& \underline{\tilde e^{(a)}{}_{\beta}} + \underline{\tilde e^{(a)}{}_{\beta, C}} \tilde{\xi}^C 
+ {1\over 2} (\underline{\tilde e^{(a)}{}_{\beta, C^1 C^2} } - \underline{\tilde{\gamma}^{\dlt}{}_{C^1C^2} \tilde e^{(a)}{}_{\beta, \dlt} } - \underline{\tilde{\gamma}^D{}_{C^1C^2} \tilde e^{(a)}{}_{\beta, D} } - \underline{\tilde e^{(a)}{}_{\dlt} \del_{\beta} \tilde{\gamma}^{\dlt}{}_{C^1C^2} } )  \tilde{\xi}^{C^1} \tilde{\xi}^{C^2} \cr 
&& + \cdots ~, \label{del-beta-prl} \\ 
&& \del_{\beta} \tilde{\exp}_{\perp}^C(x, \tilde \xi) \tilde{e}^{(a)}{}_C (\tilde{ \exp}_{\parallel}(x, \tilde \xi), \tilde{\exp}_{\perp}(x, \tilde \xi)) ~, \cr
&=& - {1\over 2} \underline{\tilde e^{(a)}{}_C \del_{\beta} \tilde{\gamma}^C{}_{C^1C^2} } \tilde{\xi}^{C^1} \tilde{\xi}^{C^2} + \cdots ~, \label{del-beta-perp} \\ 
&& \tilde{\del}_B \tilde{\exp}_{\parallel}^{\dlt} (x, \tilde \xi) \tilde{e}^{(a)}{}_{\dlt} (\tilde{ \exp}_{\parallel}(x, \tilde \xi), \tilde{\exp}_{\perp}(x, \tilde \xi)) ~, \cr
&=& - \underline{\tilde e^{(a)}{}_{\dlt} \tilde{\gamma}^{\dlt}{}_{B C}} \tilde{\xi}^C 
- (\underline{\tilde e^{(a)}{}_{\dlt, C^1} \tilde{\gamma}^{\dlt }{}_{B C^2}} 
+ {1\over 2} \underline{\tilde e^{(a)}{}_{\dlt} \tilde{\gamma}^{\dlt}{}_{B C^1C^2}} ) \tilde{\xi}^{C^1} \tilde{\xi}^{C^2} + \cdots ~, \label{del-B-prl} \\ 
&& \tilde{\del}_B \tilde{\exp}_{\perp}^C(x, \tilde \xi) \tilde{e}^{(a)}{}_C (\tilde{ \exp}_{\parallel}(x, \tilde \xi), \tilde{\exp}_{\perp}(x, \tilde \xi)) ~, \cr
&=& \underline{\tilde e^{(a)}{}_B} + (\underline{\tilde e^{(a)}{}_{B, C}} - \underline{\tilde e^{(a)}{}_D \tilde{\gamma}^D{}_{B C}} ) \tilde{\xi}^C \cr
&& + \lt\{- \underline{\tilde e^{(a)}{}_{D, C^1} \tilde{\gamma}^D{}_{B C^2} } 
+ {1\over 2}(\underline{\tilde e^{(a)}{}_{B, C^1 C^2} } - \underline{\tilde{\gamma}^{\dlt}{}_{C^1C^2} \tilde e^{(a)}{}_{B, \dlt}} - \underline{\tilde{\gamma}^D{}_{C^1C^2} \tilde e^{(a)}{}_{B, D} } \rt. \cr
&& \lt. - \underline{\tilde e^{(a)}{}_D \tilde{\gamma}^D{}_{B C^1C^2}} ) \rt\}  \tilde{\xi}^{C^1} \tilde{\xi}^{C^2} + \cdots ~. \label{del-B-perp}
\eea

Armed with the above results, we first derive eq.(\ref{calQ-beta}). Substituting (\ref{del-beta-prl}, \ref{del-beta-perp}) in (\ref{calQ-def}) one first shows,
\bea
{\cal Q}^{(a)}{}_{\beta} &=& {\cal Q}_0^{(a)}{}_{\beta} + {\cal Q}_1^{(a)}{}_{\beta} + {\cal Q}_2^{(a)}{}_{\beta} + O(\tilde \xi^3) ~, 
\eea
where,
\bea
{\cal Q}_0^{(a)}{}_{\beta} &=& \underline{\tilde e^{(a)}{}_{\beta}} ~, \label{calQ0-beta} \\
{\cal Q}_1^{(a)}{}_{\beta} &=& (\underline{\tilde e^{(a)}{}_{\beta, C}} 
- \dlt^a_D \underline{\tilde e^{(D)}{}_{C, \beta}} ) \tilde{\xi}^C ~, \cr
&=& (\underline{\tilde \gamma^D_{\beta C} \tilde e^{(a)}{}_D } - \dlt^a_A \underline{\tilde \gamma^D_{\beta C} \tilde e^{(A)}{}_D }  - \underline{\tilde \omega_C{}^{(a)}{}_{(d)} \tilde e^{(d)}{}_{\beta}} + \dlt^a_A \underline{\tilde \omega_{\beta}{}^{(A)}{}_C} ) \tilde{\xi}^C ~, \cr
&=& \lt\{ \begin{array}{ll}
0 & a = \alpha~, \cr
\underline{\tilde \omega_{\beta}{}^{(A)}{}_C} \tilde{\xi}^C & a = A ~.
\end{array} \rt.
\label{calQ1-beta} \\
{\cal Q}_2^{(a)}{}_{\beta} &=& \lt\{{1\over 2} \underline{\tilde e^{(a)}{}_{\beta, C^1 C^2} } 
- \underline{e^{(E)}{}_{C^1, \beta}  e_{(E)}{}^D} (\underline{\tilde e^{(a)}{}_{D, C^2} } 
- \underline{\tilde e^{(a)}{}_b \tilde{\gamma}^b{}_{D C^2}} ) \rt. \cr
&& \lt. - {1\over 2}(\underline{\tilde{\gamma}^{\dlt}{}_{C^1C^2} \tilde e^{(a)}{}_{\beta, \dlt} } +  \underline{\tilde{\gamma}^D{}_{C^1C^2} \tilde e^{(a)}{}_{\beta, D}} + \underline{\tilde e^{(a)}{}_c \del_{\beta} \tilde{\gamma}^c{}_{C^1C^2}} ) \rt\} \tilde{\xi}^{C^1} \tilde{\xi}^{C^2} ~, \cr
&=& {1\over 2} ( \underline{\tilde e^{(a)}{}_{\beta, C^1 C^2}} - \underline{\tilde \gamma^d{}_{C^1C^2} \tilde e^{(a)}{}_{\beta, d}} - \underline{\tilde e^{(a)}{}_d \del_{\beta} \tilde \gamma^d{}_{C^1C^2} } ) \tilde{\xi}^{C^1} \tilde{\xi}^{C^2}~, \cr
&=& {1\over 2} \underline{\tilde r^{(a)}{}_{C^1C^2\beta} } \tilde{\xi}^{C^1} \tilde{\xi}^{C^2}
\label{calQ2-beta}
\eea
The above results prove eq.(\ref{calQ-beta}). Before proceeding further we explain the computations leading to eqs.(\ref{calQ1-beta}) and (\ref{calQ2-beta}). In the second line of (\ref{calQ1-beta}) we have used,
\bea
\underline{\tilde e^{(a)}{}_{b, c} }  
&=& \underline{\tilde e^{(a)}{}_D \tilde{\gamma}^D_{b c} } - \underline{\tilde{\omega}_c{}^{(a)}{}_b}   ~,
\label{Dtot-tilde}
\eea
which is the statement of vanishing of total covariant derivative of vielbein in $\tilde z$-system and,
\bea
\underline{\tilde \gamma^{\al}_{\beta C}} &=& 0 ~, 
\eea
which is obtained by using the coordinate transformation (\ref{zb-ztilde}), under which the Christoffel symbols transform as a tensor, and the fact that $\sum_p O_{\ha p} = 0~, \forall \ha \neq 0$. In the third line of eq.(\ref{calQ1-beta}) we have used eq.(\ref{xi-omega}) and $\underline{\tilde e^{(a)}{}_B} = \dlt^a{}_A \underline{\tilde e^{(A)}{}_B}$ which can again be shown using argument similar to the above. To arive at the second line of (\ref{calQ2-beta}) we noted that,
\bea
\tilde{\xi}^{C^2} (\underline{\tilde e^{(a)}{}_{D, C^2}} -  \underline{\tilde \gamma^b{}_{D C^2} \tilde e^{(a)}{}_b} ) &=& - \tilde{\xi}^{C^2} \underline{\tilde \omg_{C^2}{}^{(a)}{}_{(b)} \tilde e^{(b)}{}_D } = 0~,
\eea
which is obtained by using (\ref{Dtot-tilde}) and (\ref{xi-omega}). Then the third line follows from the following argument. Setting $\tilde D^{tot}_{C^1} \tilde D^{tot}_{C^2} \tilde e^{(a)}{}_{\beta} = 0$ and using (\ref{xi-omega}) and (\ref{xi2-D-omega}) one first shows,
\bea
\xit^{C^1} \xit^{C^2} (\underline{\tilde e^{(a)}{}_{\beta, C^1 C^2}} - \underline{\gammat^d_{C^1C^2} \tilde e^{(a)}{}_{\beta,d}} ) 
&=& \xit^{C^1} \xit^{C^2} ( \underline{\delt_{C_1} \gammat^d_{C^2 \beta}}  + \underline{\gammat^e_{C^2 \beta} \gammat^d_{eC^1} } - \underline{\gammat^e_{C^1C^2} \gammat^d_{e\beta}}  ) 
\underline{\tilde e^{(a)}{}_d} ~. 
\label{xi2del2}
\cr &&
\eea
Then substituting the above result into the second line of (\ref{calQ2-beta}) one arrives at the third line.

We now proceed to derive eq.(\ref{calQ-B}). Using (\ref{del-B-prl}) and (\ref{del-B-perp}) in eq.(\ref{calQ-def}) and simplifying using (\ref{Dtot-tilde}, \ref{xi-omega}) one first shows,
\bea
{\cal Q}^{(a)}{}_B &=& \underline{e_{(B)}{}^D \tilde e^{(a)}{}_D} 
 + \underline{e_{(B)}{}^D \tilde{\cal U}^{(a)}{}_D} + O(\tilde \xi^3) ~, 
\eea
where,
\bea
\underline{\tilde{\cal U}}^{(a)}{}_D &=& \lt\{- \underline{\tilde e^{(a)}{}_{e, C^1} \tilde \gamma^e{}_{D C^2}} 
+ {1\over 2}( \underline{\tilde e^{(a)}{}_{D, C^1 C^2}} - \underline{\tilde \gamma^e{}_{C^1C^2} \tilde e^{(a)}{}_{D, e}} 
- \underline{\tilde e^{(a)}{}_e \tilde \gamma^e{}_{D C^1C^2}} ) \rt\}  \tilde{\xi}^{C^1} \tilde{\xi}^{C^2} ~,
\cr &&
\eea
We now massage the above expression using various tricks used so far,
\bea
\underline{\tilde{\cal U}}^{(a)}{}_D &=& 
\lt[-(\underline{\gammat^d_{C^1e} \tilde e^{(a)}{}_d} - \underline{\omgt_{C^1}{}^{(a)}{}_e} ) \underline{\gammat^e_{DC^2}} + {1\over 2} \lt\{(\underline{ \delt_{C_1} \gammat^d_{C^2 D} } + \underline{\gammat^e_{C^2 D} \gammat^d_{eC^1}} - \underline{\gammat^e_{C^1C^2} \gammat^d_{eD} } ) \underline{\tilde e^{(a)}{}_d } \rt. \rt. \cr
&& \lt. \lt. - {1\over 3} \underline{\tilde e^{(a)}{}_e } (\underline{\Dt_D \gammat^e_{C^1C^2} } + \underline{\Dt_{C^2} \gammat^e_{DC^1} } + \underline{\Dt_{C^1} \gammat^e_{C^2D} } ) \rt\} \rt] \xit^{C^1} \xit^{C^2}  ~, \cr
&=& \underline{\tilde e^{(a)}{}_d} \lt[- \underline{\gammat^d_{C^1e} \gammat^e_{DC^2}} +  {1\over 2} (\underline{\delt_{C_1} \gammat^d_{C^2 D}}  + \underline{\gammat^e_{C^2 D} \gammat^d_{eC^1} } - \underline{\gammat^e_{C^1C^2} \gammat^d_{eD}} ) \rt. \cr 
 && - {1\over 6} (\underline{\tilde{\del}_D \gammat^d_{C^1C^2}} - \underline{\gammat^e_{DC^1} \gammat^d_{eC^2}} - \underline{\gammat^e_{DC^2} \gammat^d_{eC^1} } + \underline{\tilde{\del}_{C^2} \gammat^d_{DC^1}} - \underline{\gammat^e_{C^2D} \gammat^d_{eC^1}} - \underline{\gammat^e_{C^2 C^1} \gammat^d_{eD}  } \cr
 && \lt. + \underline{\tilde{\del}_{C^1} \gammat^d_{C^2D} } - \underline{\gammat^e_{C^1C^2} \gammat^d_{eD} } - \underline{\gammat^e_{C^1D} \gammat^d_{eC^2}  } ) \rt] \xit^{C^1} \xit^{C^2}  ~, \cr
&=& {1\over 6} \underline{\et}^{(a)}{}_d \lt[\underline{\delt_{C_1} \gammat^d_{C^2 D}} -  \underline{\tilde{\del}_D \gammat^d_{C^1C^2}} +  \underline{\gammat}^d_{C^1e} \underline{\gammat}^e_{C^2 D} -  \underline{\gammat}^d_{De}  \underline{\gammat}^e_{C^1C^2}  \rt] \xit^{C^1} \xit^{C^2}  ~, \cr
&=& {1\over 6} \underline{\tilde{r}^{(a)}{}_{C^1C^2D}} \xit^{C^1} \xit^{C^2}  ~,
\eea
proving eq.(\ref{calQ-B}).

\section{Computation of metric-expansion}
\label{a:metric}

For an arbitrary submanifold-embedding $\cM \hookrightarrow \cL$, as considered in Appendix \ref{a:tubular}, the metric-expansion up to $4$-th order can be directly obtained from the results in \cite{tubular}. It is given by,
\bea
\hat g_{\al \beta} 
&=&  \underline{ G_{\al \beta} } + (\underline{\hat \omg_{\al \beta C}} + \underline{\hat \omg_{\beta \al C} } ) \hat y^C \cr
&&
+ \lt( \underline{\hat r_{\al C_1C_2 \beta} } 
+ \underline{\hat \omg_{\al}{}^a{}_{C_1} \hat \omg_{\beta a C_2} } \rt) \hat y^{C_1 C_2}   \cr
&&
+ \lt\{  {1\over 3 } \underline{ \hat D^{tot}_{C_1} \hat r_{\al C_2 C_3 \beta} } 
+ {2\over 3} (\underline{\hat r_{\al C_1C_2 a } \hat \omg_{\beta}{}^a{}_{C_3} } + \al \leftrightarrow \beta )  \rt\}  \hat y^{C_1C_2C_3}  \cr
&& 
+ \lt\{ {1 \over 12} \underline{ \hat D^{tot}_{C_1 C_2} \hat r_{\al C_3C_4 \beta} }
+ {1\over 3} \underline{\hat r_{\al C_1C_2 a } \hat r^a{}_{ C_3 C_4 \beta} }
+  {1\over 4} (\underline{ \hat D^{tot}_{C_1 } \hat r_{\al C_2 C_3 a } \hat \omg_{\beta}{}^a{}_{C_4} }  
+ \al \leftrightarrow \beta )  \rt. \cr
&& \lt. 
+ {1\over 3}  \underline{\hat r_{a C_1C_2 b} \hat \omg_{\al}{}^a{}_{C_3} \hat \omg_{\beta}{}^b{}_{C_4} } \rt\} 
\hat y^{C_1 \cdots C_4}  ~,   
\label{ghat-al-beta-gen} \\
&& \cr
\hat g_{\al B } 
&=& \underline{\hat \omg_{\al B C} } \hat y^C 
+  {2 \over 3} \underline{ \hat r_{\al C_1 C_2 B} } \hat y^{C_1 C_2 } 
+ \lt( {1 \over 4 } \underline{ \hat D^{tot}_{C_1} \hat r_{\al C_2 C_3 B} } 
+ {1 \over 3} \underline{\hat \omg_{\al }{}_{a C_1} \hat r^a{}_{ C_2 C_3 B} } \rt) \hat y^{C_1 C_2 C_3 }  \cr
&&
+ \lt( {1 \over 15} \underline{ \hat D^{tot}_{C_1 C_2 } \hat r_{\al C_3 C_4 B} } 
+ {2 \over 15 } \underline{ \hat r_{\al C_1C_2 (d)} \hat r^{(d)}{}_{C_3 C_4 B } } 
+ {1 \over 6 } \underline{\hat \omg_{\al }{}_{aC_1}  \hat D^{tot}_{C_2} \hat r^a{}_{C_3 C_4 B} } \rt) \hat y^{C_1 \cdots C_4 } ~,  \cr
&&
\label{ghat-al-B-gen} \\
\hat g_{AB} 
&=& \eta_{AB} + {1 \over 3} \underline{ \hat r_{A C_1 C_2 B} } \hat y^{C_1 C_2 } 
+ {1 \over 6 }  \underline{ \hat D^{tot}_{C_1} \hat r_{A C_2 C_3 B} } \hat y^{C_1 C_2 C_3 }  \cr
&&
+ \lt( {1 \over 20}  \underline{ \hat D^{tot}_{C_1 C_2 } \hat r_{A C_3 C_4 B} } 
+ {2 \over 45} \underline{ \hat r_{A C_1C_2 b} \hat r^b{}_{C_3 C_4 B } } \rt)  \hat y^{C_1 \cdots C_4 } ~.
\label{ghat-A-B-gen}
\eea
The coefficient of the linear term in (\ref{ghat-al-beta-gen}), namely,
\bea
\hat s_{\al \beta C} &=& \underline{\hat \omg_{\al \beta C}} + \underline{\hat \omg_{\beta \al C}} ~,  
\eea
gives the second fundamental form \cite{kobayashi-nomizu} of the submanifold embedding. Absence of this term indicates that the submanifold is totally geodesic, which is the case for all the examples that we study in this work. Notice that the terms appearing in the above expansions are of the following general form, 
\bea 
\underline{\hat u^{(n)}_{ab} } &=& \underline{\hat t_{ab C^1 \cdots C^n}} \yh^{C^1} \cdots \yh^{C^n} ~, 
\eea
where $\hat t$ can be a single tensor or product of tensors with indices contracted.

In order to obtain the tubular expansion of metric around $\Dlt \hookrightarrow \cM^N$ and 
$\Dlt \hookrightarrow \cL\cM$ in terms of $\cM$-data, all we need to do is to take the above results, specialise to these two cases and use the right Jacobian matrices to express the expansion coefficients in terms of $\cM$-data. We perform this procedure below for the two cases in order. 

\subsection{Specialisation to $\Dlt \hookrightarrow \cM^N$}
\label{sa:M^N}

In this case using the results of \S \ref{tub:M^N} we may write, when $\hat t$ is a single tensor,
\bea
\underline{\hat u^{(n)}_{ab}}  &=& \underline{K^{\bar a}{}_a K^{\bar b}{}_b K^{\bar c^1}{}_{C^1}  \cdots 
K^{\bar c^n}{}_{C^n} \bar t_{\bar a \bar b \bar c^1 \cdots \bar c^n}  } \hat y^{C^1} \cdots \hat y^{C^n} ~, \cr
&=& N^{n+w\over 2} \sum_{p, \hc^1 \cdots \hc^n} \underline{K^{\al_p }{}_a K^{\beta_p }{}_b } \{ O^T_{p \hc^1} \cdots O^T_{p \hc^n} \} T_{\al_p \beta_p (\hat \xi^1 \cdots \hat \xi^n )}(x) \hat y_{\hc^1}^{\hat \xi^1} \cdots \hat y_{\hc^n}^{\hat \xi^n} ~,
\label{form1-yhat}
\eea
which implies,
\bea
\underline{\hat u^{(n)}_{\al \beta}}  &=& N^{1+{n+w\over 2}} T_{\al \beta (\hat \xi^1 \cdots \hat \xi^n )}(x) \sum_{p, \hc^1 \cdots \hc^n} \{ (O^T_{p0})^2 O^T_{p \hc^1} \cdots O^T_{p \hc^n} \} \hat y_{\hc^1}^{\hat \xi^1} \cdots \hat y_{\hc^n}^{\hat \xi^n} ~, 
\label{uhatn-al-bet}
\\
\underline{\hat u^{(n)}_{\al B}} 
&=& N^{1+{n+w\over 2}} T_{\al (\hat \beta \hat \xi^1 \cdots \hat \xi^n) } (x) \sum_{p, \hc^1, \cdots , \hc^n} \{ O^T_{p0} O^T_{p \hb}  O^T_{p \hc^1} \cdots O^T_{p \hc^n} \}
\hat y_{\hc^1}^{\hat \xi^1} \cdots \hat y_{\hc^n}^{\hat \xi^n} ~,  
\label{uhatn-al-B}
\\
\underline{\hat u^{(n)}_{A B} }  &=& N^{1+{n+w\over 2}} T_{ (\hat \al \hat \beta \hat \xi^1 \cdots \hat \xi^n )} (x) 
\sum_{p, \hc^1, \cdots , \hc^n} \{ O^T_{p \ha} O^T_{p \hb} O^T_{p \hc^1} \cdots O^T_{p \hc^n} \} \hat y_{\hc^1}^{\hat \xi^1} \cdots \hat y_{\hc^n}^{\hat \xi^n} ~.
\label{uhatn-A-B}
\eea

The same general forms of the results hold true when $\hat t$ is a product of tensors as well. In that case the tensor $T$ in $\cM$ should also be a product of the corresponding tensors. For example,
\bea
\underline{\hat t^1_{\alpha C^1\cdots C^{n} e} \hat t^{2 e}{}_{\beta D^1 \cdots D^{m}}} \to T^1_{\alpha (\hat \xi^1 \cdots \hat \xi^{n}) \eta} (x) T^{2 \eta}{}_{\beta (\hat \dlt^1 \cdots \hat \dlt^{m})} (x) ~,
\eea
where $T^1$ and $T^2$ correspond to $\hat t^1$ and $\hat t^2$ respectively.

The results (\ref{uhatn-al-bet}, \ref{uhatn-al-B}) and (\ref{uhatn-A-B}), when written in terms of $\xi_p$, take the following forms,
\bea
\underline{\hat u^{(n)}_{\al \beta}}  &=& N^{1+{w\over 2}} T_{\al \beta \gamma^1 \cdots \gamma^n }(x) \sum_p (O^T_{p0})^2 \xi_p^{\gamma^1 \cdots \gamma^n} ~, 
\label{uhatn-al-bet-xi}
\\
\underline{\hat u^{(n)}_{\al B}} 
&=& N^{1+{w\over 2}} T_{\al (\hat \beta) \gamma^1 \cdots \gamma^n } (x) \sum_p O^T_{p0} O^T_{p \hb}
\xi_p^{\gamma^1 \cdots \gamma^n} ~,  
\label{uhatn-al-B-xi}
\\
\underline{\hat u^{(n)}_{AB} }  &=& N^{1+{w\over 2}} 
T_{ (\hat \al \hat \beta) \gamma^1 \cdots \gamma^n} (x) 
\sum_p O^T_{p \ha} O^T_{p \hb} \xi_p^{\gamma^1 \cdots \gamma^n} ~.
\label{uhatn-A-B-xi}
\eea
It is now easy to see from eq.(\ref{uhatn-al-bet-xi}) that $\underline{\hat u_{\al \beta}^{(1)}}$ vanishes due to the fact that $O_{0p}$ is independent of $p$ and the condition in eq.(\ref{xi-transverse}). Similar argument can also be given using eq.(\ref{uhatn-al-bet}). This shows that $\Dlt \hookrightarrow \cM^N$ is totally geodesic. Below we compute the metric-expansion for $N=2, 3$. All the tensors appearing in this expansion has Weyl weight $w=-2$.

\noindent
\underline{$N=2$}

In this case there is only one set of transverse coordinates, namely $\hat y_1 = \hat y$. We take the following $SO(2)$ matrix,
\bea
O &=& {1\over \sqrt{2}} \pmatrix{ 1 & 1 \cr -1 & 1 } ~. 
\label{SO2}
\eea
Equations (\ref{uhatn-al-bet}, \ref{uhatn-al-B}) and (\ref{uhatn-A-B}) take the following forms (for $w=-2$),
\bea
\underline{\hat u^{(n)}_{\al \beta}}  &=& T_{\al \beta (\hat \xi^1 \cdots \hat \xi^n )}(x) 
{1\over 2} [1+ (-1)^n] \hat y^{\hat \xi^1 \cdots \hat \xi^n} ~ , \cr
\underline{\hat u^{(n)}_{\al \hat \beta}} 
&=& T_{\al (\hat \beta \hat \xi^1 \cdots \hat \xi^n) } (x) {1\over 2} [1-(-1)^n] \hat y^{\hat \xi^1 \cdots \hat \xi^n} ~, 
\cr
\underline{\hat u^{(n)}_{\hat \al \hat \beta} }  
&=& T_{ (\hat \al \hat \beta \hat \xi^1 \cdots \hat \xi^n )} (x) 
{1\over 2} [1+(-1)^n] \hat y^{\hat \xi^1 \cdots \hat \xi^n} ~.
\eea
The final results for the metric-expansion, obtained using the above equations, are given in eqs.(\ref{g-al-beta-2}, \ref{g-al-B-2}) and (\ref{g-A-B-2}).

\noindent
\underline{$N=3$}

In this case we describe the results in terms of $\xi_p$ using eqs.(\ref{uhatn-al-bet-xi}, \ref{uhatn-al-B-xi}) and (\ref{uhatn-A-B-xi}). The $SO(3)$ matrix is taken to be,
\bea
O &=& \pmatrix{ {1\over \sqrt{3}} & {1\over \sqrt{3}} & {1\over \sqrt{3}} \cr
- {1\over \sqrt{2}} & {1\over \sqrt{2}} & 0 \cr
- {1\over \sqrt{6}} & - {1\over \sqrt{6}} & \sqrt{2\over 3} } ~.
\label{SO3}
\eea
This gives the following results,
\bea
\underline{\hat u^{(n)}_{\al \beta}}  &=& {1\over 3} T_{\al \beta \gamma^1 \cdots \gamma^n } (x) \sum_p \xi_p^{\gamma^1 \cdots \gamma^n} ~, \cr
\underline{\hat u^{(n)}_{\al \hat \beta_1}} 
&=& T_{\al (\hat \beta_1) \gamma^1 \cdots \gamma^n } (x) \sum_p O^T_{p0} O^T_{p 1}
\xi_p^{\gamma^1 \cdots \gamma^n} = {1\over \sqrt{6}} T_{\al (\hat \beta_1) \gamma^1 \cdots \gamma^n } (x)  
(-\xi_1^{\gamma^1 \cdots \gamma^n} + \xi_2^{\gamma^1 \cdots \gamma^n} ) ~,  \cr 
\underline{\hat u^{(n)}_{\al \hat \beta_2}}  &=& {1\over 3\sqrt{2}} T_{\al (\hat \beta_2) \gamma^1 \cdots \gamma^n } (x) ( -\xi_1^{\gamma^1 \cdots \gamma^n} - \xi_2^{\gamma^1 \cdots \gamma^n} + 2 \xi_3^{\gamma^1 \cdots \gamma^n}) ~, \cr
\underline{\hat u^{(n)}_{\hat \al_1 \hat \beta_1} }  &=& 
\lt\{ \begin{array}{ll}
T_{(\hat \al_1 \hat \beta_1)}(x) & n = 0 ~, \cr
\displaystyle{{1 \over 2} T_{ (\hat \al_1 \hat \beta_1) \gamma^1 \cdots \gamma^n} (x) 
(\xi_1^{\gamma^1 \cdots \gamma^n} + \xi_2^{\gamma^1 \cdots \gamma^n} ) } & n \neq 0 ~,
\end{array} \rt. \cr
\underline{\hat u^{(n)}_{\hat \al_1 \hat \beta_2} }  &=& 
T_{ (\hat \al_1 \hat \beta_2) \gamma^1 \cdots \gamma^n} (x) 
\sum_p O^T_{p 1} O^T_{p 2} \xi_p^{\gamma^1} \cdots \xi_p^{\gamma^n} 
= {1 \over 2 \sqrt{3}} 
T_{ (\hat \al_1 \hat \beta_2) \gamma^1 \cdots \gamma^n} (x) 
( \xi_1^{\gamma^1 \cdots \gamma^n} - \xi_2^{\gamma^1 \cdots \gamma^n} ) ~, \cr 
\underline{\hat u^{(n)}_{\hat \al_2 \hat \beta_2} } 
&=& \lt\{ \begin{array}{ll}
T_{(\hat \al_2 \hat \beta_2)}(x) & n=0 ~, \cr 
\displaystyle{{ 1 \over 6} T_{ (\hat \al_2 \hat \beta_2) \gamma^1 \cdots \gamma^n} (x) 
( \xi_1^{\gamma^1 \cdots \gamma^n} + \xi_2^{\gamma^1 \cdots \gamma^n} + 4 \xi_3^{\gamma^1 \cdots \gamma^n} ) } & n \neq 0 
\end{array} \rt. ~.
\eea
The metric-expansion, obtained by using the above results, are given in eqs.(\ref{g-al-beta-3}, \ref{g-al-betahat1-3}, \ref{g-al-betahat2-3}, \ref{g-alhat1-betahat1-3}, \ref{g-alhat1-betahat2-3}) and (\ref{g-alhat2-betahat2-3}).

\subsection{Specialisation to $\Dlt \hookrightarrow \cL \cM$}
\label{sa:LM}

In this case the results can be obtained simply by taking the results of \S \ref{sa:M^N} (in $\xi$-form), replacing $N$ by $m=(2N+1)$, $O$ by $U$ (as given in eqs.(\ref{U-def})), $O^T$ by $U^{\dagger}$, and taking the large-$N$ limit. This leads to the following equations for $w=-2$,
\bea
\underline{\hat u^{(n)}_{\al \beta}}  &=& T_{\al \beta \gamma^1 \cdots \gamma^n }(x) 
\sum_p (U^*_{0p})^2 \xi_p^{\gamma^1 \cdots \gamma^n} = T_{\al \beta \gamma^1 \cdots \gamma^n }(x) 
{1\over m} \sum_p  \xi_p^{\gamma^1 \cdots \gamma^n} 
~, \cr
&\to &  T_{\al \beta (\hat \xi^1 \cdots \hat \xi^n) }(x) \oint {d\s \over 2\pi} \hat Y^{\hat \xi^1 \cdots \hat \xi^n}(\s) ~,
\label{uhatn-al-bet-LM} 
\\ && \cr
\underline{\hat u^{(n)}_{\al B}} 
&=& T_{\al (\hat \beta) \gamma^1 \cdots \gamma^n } (x) \sum_p U^*_{0p} U^*_{\hb p}  \xi_p^{\gamma^1 \cdots \gamma^n} = T_{\al (\hat \beta) \gamma^1 \cdots \gamma^n } (x) {1\over m} \sum_p e^{i\hb \s}  \xi_p^{\gamma^1 \cdots \gamma^n} ~,  \cr
&\to &  T_{\al (\hat \beta \hat \xi^1 \cdots \hat \xi^n ) } (x) \oint {d\s \over 2\pi} e^{i\hb \s} \hat Y^{\hat \xi^1 \cdots \hat \xi^n}(\s) ~,
\label{uhatn-al-B-LM}
\\
\underline{\hat u^{(n)}_{A B} }  &=& T_{ (\hat \al \hat \beta) \gamma^1 \cdots \gamma^n } (x) 
\sum_p U^*_{\ha p} U^*_{\hb p} \xi_p^{\gamma^1 \cdots \gamma^n}
= T_{(\hat \al \hat \beta) \gamma^1 \cdots \gamma^n } (x) {1\over m} \sum_p e^{i(\ha + \hb) \s}  \xi_p^{\gamma^1 \cdots \gamma^n} ~, \cr
&\to & T_{(\hat \al \hat \beta) \gamma^1 \cdots \gamma^n } (x) \oint {d\s \over 2\pi}  e^{i(\ha + \hb) \s} \hat Y^{\hat \xi^1 \cdots \hat \xi^n}(\s) ~,
\label{uhatn-A-B-LM}
\eea
where an $\to$ indicates taking the large-$N$ limit. Using the above results one finds the metric-expansion which is given in eqs.(\ref{g-al-bet-LM}, \ref{g-al-B-LM}) and (\ref{g-A-B-LM}).

\section{Construction of ``Killing vector" at finite $N$}
\label{a:killing-finite}

Here we will perform the following steps,
\begin{enumerate}
\item
Construct a vector field in ${\cal L}^{(N)}\cM$ in DPC such that its large-$N$ limit taken following \S \ref{s:LM-DPC} is given by (\ref{kappa-sigma}), 
\item
Perform the coordinate transformation to find the vector field in FNC.
\item
Take large-$N$ limit back to arrive at (\ref{kappah}).
\end{enumerate}

To perform the first step we first consider an $m=(2N+1)$-point configuration in $\cM$ given by the coordinates $x_p$ ($p=1, \cdots , m$), and then go to the tangent space $T_x \cM$ where $x \in \cM$ is the CM. Recall that this configuration is given by a set of $m$ vectors in $T_x\cM$, namely $\xi_p^{\al}$ satisfying (\ref{xi-transverse}). We now consider the following convex combination of $\xi_p$'s,
\bea
\xi^{\al}(\lambda_1 \cdots \lambda_m) &=& \sum_{p=1}^m \lambda_p \xi_p^{\al} ~, \quad \lambda_p \geq 0 ~, \hbox{ and } \sum_{p=1}^m \lambda_p = 1 ~,
\eea
which defines a polygon in $T_x \cM$. This, in turn, defines a polygon in $\cM$ with $x_p$ as vertices through exponential map,
\bea
x^{\al}(\lambda_1, \cdots \lambda_m) = x^{\al} + \Exp_x^{\al}(\xi(\lambda_1, \cdots , \lambda_m)) ~.
\eea 
Therefore the $p$-th vertex is obtained by choosing $\lambda_p=1~, \lambda_{q\neq p} = 0$. Note that this is not a geodesic polygon. Why we choose to consider this tangent-space-polygon and not the geodesic one will be clarified later.
Because of cyclic ordering the polygon is oriented (from lower value of $p$ to higher value) and the segment $x_p x_{p+1}$ can be written as,
\bea
x^{\al}(\beta_p) &=& x^{\al} + \Exp_x^{\al}((1-\beta_p)\xi_p + \beta_p \xi_{p+1}) ~, \quad 0 \leq \beta_p \leq 1 ~.
\eea
Notice that cyclicity implies: $x_{m+1}=x_1$ and $\xi_{m+1}=\xi_1$. The tangent vector to the above segment at the $p$-th vertex along the direction of its orientation is given by,
\bea
\Dlt_p^{\al} (x_p) &\equiv& \lt({\del x^{\al}(\beta_p) \over \del \beta_p}\rt)_{\beta_p=0} = \Dlt \xi_p^{\gamma}
\del_{\zeta^{\gamma}} \Exp_x^{\al}(\zeta)|_{\zeta = \xi_p} ~,  
\label{Dlt-p}
\eea
where $\Dlt \xi_p^{\gamma} = (\xi^{\gamma}_{p+1}-\xi^{\gamma}_p)$. Our desired vector field in ${\cal L}^{(N)}\cM$ is given, in DPC, by,
\bea
\bar \Dlt^{\bar a} &=& (\Dlt_1^{\al_1}(x_1), \Dlt_2^{\al_2}(x_2), \cdots , \Dlt_m^{\al_m}(x_m) )~.
\label{Dlt-bar}
\eea
We now discuss the large-$N$ limit. To this end let us consider two vertices given by $\xi_{p-{\dlt p\over 2}}$ and $\xi_{p+{\dlt p\over 2}}$, where $\dlt p$ is a suitable range such that ${\dlt p \over m}$ is infinitesimally small at large $N$. The segment between these two vertices is given by,
\bea
\xi^{\al}(s_p) &=& (1-s_p)\xi_{p-{\dlt p\over 2}} + s_p \xi_{p+{\dlt p\over 2}} ~, \quad 0 \leq s_p \leq 1~.
\label{xi-sp}
\eea
In general this segment is not on our actual polygon in $T_x\cM$. However, at large $N$ (when ${\dlt p \over m}$ is infinitesimally small) and the loop considered is a smooth one any intermediate vertex, i.e. $\xi_q$ for $(p-{\dlt p\over 2}) < q < (p+{\dlt p \over 2})$
sits on the above segment. Given this, we now define the large-$N$ analogue of $\Dlt_p(x_p)$ in (\ref{Dlt-p}),
\bea
\kappa^{\al}_p(x_p) &=& 2\pi {(\xi^{\gamma}_{p+{\dlt p\over 2}} - \xi^{\gamma}_{p-{\dlt p\over 2}}) \over {\dlt p \over m}} \del_{\zeta^{\gamma}} \Exp^{\al}_x(\zeta)|_{\zeta = \xi_p} \to \kappa^{\al}(\s) = \del \xi^{\gamma}(\s) \del_{\zeta^{\gamma}} \Exp^{\al}_x(\zeta)|_{\zeta = \xi(\s)}~, \cr &&
\label{kappa-large-N}
 \eea 
where the arrow indicates the large-$N$ limit. Finally, going back to the first line of (\ref{xp-xip}) and interpreting it in the large-$N$ limit one concludes that $\kappa^{\al}(\s)$, as given above, is same as that in eq.(\ref{kappa-sigma}). 

We now perform the second step. Notice that because of cyclicity, $\Dlt \xi_p$ satisfy the following condition\footnote{Equation (\ref{Dlt-xi-transverse}), which is a transversality condition, will not be satisfied if we had considered a geodesic polygon instead of the tangent-space-polygon.},
\bea
\sum_p \dlt^{\al}{}_{\al_p} \Dlt \xi_p^{\al_p} &=& 0~,
\label{Dlt-xi-transverse}
\eea
Equation (\ref{Dlt-p}), along with the above condition implies that the vector field in (\ref{Dlt-bar}), when expressed in FNC, takes the following form, 
\bea
\hat \Dlt^a &=& (0, \Dlt \hat y^A) ~,
\label{Dlt-hat}
\eea
which may be taken to be the definition of a vector field being transverse to the submanifold. We show this below in the reverse direction using the arguments of Appendices \ref{sa:arbit-TGC} and \ref{sa:CT},
\bea
\Dlt_p^{\al_p} &=& \sqrt{m} (R^{\dagger})^{\bar a}{}_a  \tilde\Dlt^a ~, \cr
&=& \sqrt{m} (R^{\dagger})^{\bar a}{}_a \lt( \del \tilde z^a \over \del \xi^B \rt) {1\over \sqrt{m}} R^B{}_{\bar d} \Dlt \bar \xi^{\bar d}  ~, \quad [\hbox{$\tilde \Dlt$ is transverse to submanifold}]\cr
&=& (R^{\dagger})^{\bar e}{}_B {\del \over \del \bar \xi^{\bar e}}  \lt[ \xi_p^{\alpha_p} - \sum_{n \geq 0} {1\over (n+2)!}
\Gamma^{\alpha_p}{}_{\beta_p^1 \cdots \beta_p^{n+2}}(x) \xi_p^{\beta_p^1} \cdots \xi_p^{\beta_p^{n+2}} \rt] R^B{}_{\bar d} \Dlt \bar \xi^{\bar d} ~, \cr
&=& {\del \over \del \bar \xi^{\bar d}}  \lt[ \xi_p^{\alpha_p} - \sum_{n \geq 0} {1\over (n+2)!}
\Gamma^{\alpha_p}{}_{\beta_p^1 \cdots \beta_p^{n+2}}(x) \xi_p^{\beta_p^1} \cdots \xi_p^{\beta_p^{n+2}} \rt] \Dlt \bar \xi^{\bar d} ~, \quad [\hbox{Using eq.(\ref{Dlt-xi-transverse})} ] ~, \cr && \cr
&=& \Dlt \xi_p^{\gamma} \del_{\zeta^{\gamma}} \Exp_x^{\al_p}(\zeta)|_{\zeta = \xi_p}~. \quad (\hbox{Q.E.D.})
\eea
It is now straightforward to relate $\Dlt \hat y$ in (\ref{Dlt-hat}) and $\Dlt \xi_p$ in (\ref{Dlt-p}),
\bea
\Dlt \hat y^A &=& \underline{e^{(A)}{}_B} \Dlt y'^B = \underline{e^{(A)}{}_B} \Dlt \xi^B  = {1\over \sqrt{m}}  E^{(\hat \al)}{}_{\beta} (x) \sum_p U_{\ha p} \Dlt \xi_p^{\beta} ~.
\eea

The final step is to take the large-$N$ limit of the above equation. This can be done following the same procedure that we described earlier for the case of DPC. By simply comparing (\ref{Dlt-p}) and (\ref{kappa-large-N}) one concludes that in FNC the Killing vector in (\ref{kappa-large-N}) is given by eq.(\ref{kappa-cal}) in the continuum limit.

\section{Verification of Killing equation}
\label{a:killing-verification}

Here we will verify the Killing equation in vielbein form as given in (\ref{killing}). Using (\ref{kappah}) in (\ref{killing}) we first rewrite the equation in the following form,
\bea
\hat d (\eh^{(a)}{}_{\beta} ) &=& \chi^{(a)}{}_{(d)} \eh^{(d)}{}_{\beta}~, \label{killing-beta} \\
\hat d (\eh^{(a)}{}_B ) + i\hb \eh^{(a)}{}_B &=& \chi^{(a)}{}_{(d)} \eh^{(d)}{}_B ~. \label{killing-B}  
\eea
where $\hat d \equiv \displaystyle{\sum_{\hc \neq 0} i \hc \hat y^C \hat \del_C} $. These equations can be easily shown to hold true by using the following identities, 
\bea
\hat d (\eh_0^{(\alpha)}{}_{\beta} ) &=& 0 ~, \quad \hat d (\eh_0^{(a)}{}_B ) = 0 ~, 
\label{ehat0-easy} \\
\hat d (\eh_0^{(A)}{}_{\beta} )  &=&  i \ha \eh_0^{(A)}{}_{\beta} ~, 
\label{ehat0-neasy} \\
\hat d [\underline{\hat \pi^{(a)}{}_{(b)}}(\{s\}_n, \hat y)] &=& i(\ha - \hb) [\underline{ \hat \pi^{(a)}{}_{(b)}} (\{s\}_n , \hat y)] ~, \quad \ha, \hb \in \ZZ ~,
\label{dhat-pihat}
\eea
with,
\bea
\chi^{(a)}{}_{(b)} &=& i \ha \dlt^a{}_b~.
\eea
This implies,
\bea
\chi_{(a b)} &=& i \hb \eta_{ab} = i \hb \eta_{\hat \alpha \hat \beta} \dlt_{\ha + \hb, 0}~, \quad \ha, \hb \in \ZZ ~,
\eea
satisfying the condition in (\ref{chi}) as required. Notice that in the above equation we have allowed $\ha, \hb$ to take the value $0$, as in eq.(\ref{gen-coeff}). Therefore we follow the convention for index notation as explained below that equation.

Therefore all we have to do here is to prove eqs.(\ref{ehat0-easy}, \ref{ehat0-neasy}) and (\ref{dhat-pihat}). While the first two equations in (\ref{ehat0-easy}) follow trivially from (\ref{ehat0-alpha}) and the last equation in (\ref{underline-e}), one needs to do more work to prove the other ones. Below we first show eq.(\ref{ehat0-neasy}), 
\bea
\hat d (\eh_0^{(A)}{}_{\beta} )  &=& \sum_{\hc \neq 0} i \hc \underline{\omgh_{\beta}{}^{(A)}{}_C } \yh^C  ~, \cr
&=& \sum_{\hc \neq 0} i \hc R^A{}_{\bar a}\underline{K^{\bar b}{}_{\beta} K^{\bar c}{}_C \bar \omg_{\bar b}{}^{(\bar a)}{}_{\bar c} J^C{}_{\bar e} } \bar \xi^{\bar e} \cr
&=& \sum_{p, q} \sum_{\hc \neq 0} i \hc (U_{\ha p} \dlt^{\hat \al}{}_{\al_p} ) (\sqrt{N} U^*_{0p} \dlt^{\beta_p}{}_{\beta} ) (\sqrt{N} U^*_{\hc p} E_{(\hat \xi)}{}^{\xi_p}(x) ) ({1\over \sqrt{N}} \Omg_{\beta_p}{}^{(\al_p)}{}_{\xi_p}(x) ) \cr
&& ({1\over \sqrt{N}} E^{(\hat \xi)}{}_{\eta_q} (x) U_{\hc q}) \xi_q^{\eta_q} ~, \cr
&=& i \ha \Omg_{\beta}{}^{(\hat \al)}{}_{\eta }(x) {1\over \sqrt{N}} \sum_q  U_{\ha q}\xi_q^{\eta }  \cr 
&=& i \ha \eh_0^{(A)}{}_{\beta} ~, \quad (\hbox{Q.E.D.})
\eea

To show eq.(\ref{dhat-pihat}), one first notices, following the general analysis of \S \ref{s:exp-vielbein}, that,
\bea
[\underline{(\hat y .\delh )^s \hat \rho (\hat y)}]^{(a)}{}_{(b)} &=& \sum_p U_{\ha p} U^*_{p\hb} (\xi_p.\nabla)^s [{\cal R}(\xi_p;x)]^{(\hat \al)}{}_{(\hat \beta)} ~, \cr
&\to & \oint {d\s \over 2\pi} e^{-i(\ha -\hb) \s} (\xi(\s). \nabla)^s [{\cal R}(\xi(\s); x)]^{(\hat \al)}{}_{(\hat \beta)}~, \quad \ha, \hb \in \ZZ ~, \cr &&
\label{element}
\eea
where the arrow indicates continuum limit. Next, from (\ref{pi-hat}) we can write,
\bea
\hat d [\underline{(\hat y .\delh )^s \hat \rho (\hat y)}]^{(a)}{}_{(b)} 
&=& \hat d [ \hat y^{A_1} \cdots \hat y^{A_s} \hat y^{D} \hat y^{E} \underline{\hat \del_{A_1} \cdots \hat \del_{A_s} \hat r^{(a)}{}_{DE (b)}} ] ~, \cr
&=& \sum_{\ha^1 \neq 0} (i \ha^1 \hat y^{A_1} ) \hat y^{A_2} \cdots \hat y^{A_s} \hat y^{D} \hat y^{E} \underline{\hat \del_{A_1} \cdots \hat \del_{A_s} \hat r^{(a)}{}_{DE (b)} } \cr
&& + \cdots \sum_{\ha^s \neq 0} \hat y^{A_1} \cdots \hat y^{A_{s-1}} (i \ha^s \hat y^{A_s} ) \hat y^{D} \hat y^{E} \underline{\hat \del_{A_1} \cdots \hat \del_{A_s} \hat r^{(a)}{}_{DE (b)} } \cr
&& + \sum_{\hd \neq 0} \hat y^{A_1} \hat y^{A_2} \cdots \hat y^{A_s} (i \hd  \hat y^{D} ) \hat y^{E} \underline{\hat \del_{A_1} \cdots \hat \del_{A_s} \hat r^{(a)}{}_{DE (b)} } \cr
&& + \sum_{\he \neq 0} \hat y^{A_1} \hat y^{A_2} \cdots \hat y^{A_s}  \hat y^{D} (i \he \hat y^{E}) \underline{\hat \del_{A_1} \cdots \hat \del_{A_s} \hat r^{(a)}{}_{DE (b)} } ~.
\label{dhat-element}
\eea
Therefore the general form of the term to be calculated is (for $w=0$),
\bea
&& \sum_{\hd \neq 0} \underline{\hat t^{(a)}{}_{(b) D D^1 \cdots D^n } } (i \hd \hat y^D) \hat y^{D^1} \cdots \hat y^{D^n} \cr
&=& \sum_{\hd \neq 0} R^a{}_{\bar a} (R^{\dagger})^{\bar b}{}_b \underline{\bar t^{(\bar a)}{}_{(\bar b) \bar d \bar d^1 \cdots \bar d^n} 
(i\hd K^{\bar d}{}_D J^D{}_{\bar e}) (K^{\bar d^1}{}_{D^1} J^{D^1}{}_{\bar e^1}) \cdots (K^{\bar d^n}{}_{D^n} J^{D^n}{}_{\bar e^n}) } \bar \xi^{\bar e} \bar \xi^{\bar e^1} \cdots \bar \xi^{\bar e^n} ~, \cr 
&=& T^{(\hat \al)}{}_{(\hat \beta) \dlt \dlt^1 \cdots \dlt^n} (x) \sum_{\hd \neq 0, p, q} U_{\ha p} U^*_{\hb p} (i\hd U^*_{\hd p} U_{\hd q} \xi_q^{\dlt} )  \xi_p^{\dlt^1} \cdots \xi_p^{\dlt^n} ~, \cr
&\to & T^{(\hat \al)}{}_{(\hat \beta) \dlt \dlt^1 \cdots \dlt^n} (x) \sum_{\hd \neq 0} \oint {d\s \over 2\pi} {d\s' \over 2\pi}
e^{-i(\ha - \hb) \s }  e^{i\hd (\s - \s' )} \del' \xi^{\dlt}(\s') \xi^{\dlt^1}(\s) \cdots \xi^{\dlt^n}(\s) ~, \cr
&= & T^{(\hat \al)}{}_{(\hat \beta) \dlt \dlt^1 \cdots \dlt^n} (x) \oint {d\s \over 2\pi}
e^{-i(\ha - \hb) \s }  \del \xi^{\dlt}(\s) \xi^{\dlt^1}(\s) \cdots \xi^{\dlt^n}(\s) ~, \quad \ha, \hb \in \ZZ~.
\eea
Using this result and (\ref{element}) in eq.(\ref{dhat-element}) one gets,
\bea
\hat d [\underline{(\hat y .\delh )^s \hat \rho (\hat y)}]^{(a)}{}_{(b)} 
&=& \oint {d\s \over 2\pi} e^{-i(\ha -\hb) \s} \del \lt\{(\xi(\s). \nabla)^s [{\cal R}(\xi(\s); x)]^{(\hat \al)}{}_{(\hat \beta)} \rt\} ~, \cr
&=& (\ha - \hb) [\underline{(\hat y .\delh )^s \hat \rho (\hat y)}]^{(a)}{}_{(b)} ~, \quad \ha, \hb \in \ZZ~,
\eea
which, in turn, is used in the first equation in (\ref{pi-hat}) to establish the result (\ref{dhat-pihat}). With this we conclude that the reparametrization isometry is admitted by our large-$N$ geometry.

\end{document}